\documentclass{jpp}
\usepackage{graphicx}

\usepackage[utf8]{inputenc}
\usepackage[T1]{fontenc}
\usepackage{amsmath}
\usepackage{amssymb}
\usepackage{hyperref}
\usepackage{commath}
\usepackage{upgreek}
\usepackage{cancel}
\usepackage{subcaption}
\usepackage{xcolor}

\usepackage[normalem]{ulem} % for sout;
\usepackage{soul} % use \st{text} to strike-through text;

\newcommand{\normvec}{\ensuremath{\boldsymbol{\hat{\mathrm{n}}}}}
\newcommand{\normpert}{\ensuremath{\delta\boldsymbol{\mathrm{x}}\cdot \normvec}}
\newcommand*\diff{\mathop{}\!\mathrm{d}}

\shorttitle{Adjoint methods for quasisymmetry of vacuum fields on a surface}
\shortauthor{R. Nies, E. J. Paul, S. R. Hudson and A. Bhattacharjee}

\title{Adjoint methods for quasisymmetry of vacuum fields on a surface}

\author{Richard Nies\aff{1,2}
  \corresp{\email{rnies@pppl.gov}},
  Elizabeth J. Paul\aff{1,2},
  Stuart R. Hudson\aff{2}
 \and Amitava Bhattacharjee\aff{1,2}}

\affiliation{\aff{1}Department of Astrophysical Sciences, Princeton University, Princeton, NJ, 08543
\aff{2}Princeton Plasma Physics Laboratory, Princeton, NJ, 08540}

\begin{document}

\maketitle

\begin{abstract}
    Adjoint methods can speed up stellarator optimisation by providing gradient information more efficiently compared to finite-difference evaluations. Adjoint methods are herein applied to vacuum magnetic fields, with objective functions targeting quasisymmetry and a rotational transform value on a surface. To measure quasisymmetry, a novel way of evaluating approximate flux coordinates on a single flux surface  without the assumption of a neighbourhood of flux surfaces is proposed. The shape gradients obtained from the adjoint formalism are evaluated numerically and verified against finite-difference evaluations.
\end{abstract}

\section{Introduction}

The stellarator concept \citep{spitzerStellaratorConcept1958} offers a path to a steady-state and disruption-free fusion reactor with low recirculating power, but its complex three-dimensional geometry must be carefully designed to guarantee good plasma properties. In particular, the stellarator does not generally guarantee confinement of particles on collisionless trajectories due to its lack of continuous symmetry, leading to large neoclassical transport \citep{helanderTheoryPlasmaConfinement2014}. However, the use of numerical optimisation techniques has led to advanced stellarator designs with good confinement properties, culminating in the design and construction of the HSX \citep{andersonHelicallySymmetricExperiment1995} and W7-X stellarators \citep{beidlerPhysicsEngineeringDesign1990}.

Although gradient-based optimisation algorithms are generally more efficient than gradient-free algorithms, because of the large number of parameters (e.g. to represent the plasma boundary) they can be prohibitively expensive computationally if the gradients are evaluated via finite-differences. A more efficient way of obtaining gradient information is provided by adjoint methods, which were recently introduced in the stellarator optimisation field and have already found widespread application \citep{landremanComputingLocalSensitivity2018, paulAdjointMethodGradientbased2018, antonsenAdjointApproachCalculating2019, paulAdjointMethodNeoclassical2019, paulAdjointApproachCalculating2020, paulAdjointMethodsStellarator2020, paulGradientbasedOptimization3D2021, geraldiniAdjointMethodDetermining2021, giulianiSinglestageGradientbasedStellarator2020}.

Previous work \citep{antonsenAdjointApproachCalculating2019, paulAdjointApproachCalculating2020, paulGradientbasedOptimization3D2021} applied adjoint methods to ideal magnetohydrostatic (MHS) equilibria, building in the assumption of integrability, i.e. the existence of a set of nested flux surfaces. However, three-dimensional magnetic fields are generally not integrable due to the lack of continuous symmetry. Moreover, singularities arise at rational surfaces for linearised ideal MHS equilibria, making the computation of derivatives challenging \citep{paulGradientbasedOptimization3D2021}. To overcome these challenges, different equilibrium models can be considered, such as vacuum or force-free fields. We herein apply adjoint methods to vacuum magnetic fields, relinquishing the assumption of global integrability, and avoiding the singular behaviour of MHS equilibria. Modeling the plasma magnetic field as a vacuum field is justified in the limit of vanishing plasma current and $\beta$, the ratio of thermal pressure to magnetic pressure. Vacuum fields are thus broadly relevant for stellarators configurations, which tend to operate at low $\beta$ and low plasma current, as non-axisymmetric shaping of the coils is used to generate rotational transform. Moreover, optimised vacuum solutions can serve as useful starting points for the optimisation of finite-pressure equilibria \citep{boozerCurlfreeMagneticFields2019}.

We shall consider two objective functions, one targeting a rotational transform value on the boundary, and another targeting quasisymmetry on the boundary. As a subset of the larger class of omnigenous fields \citep{hallThreedimensionalEquilibriumAnisotropic1975}, for which particles are confined on collisionless trajectories, quasisymmetric fields \citep{nuhrenbergQuasihelicallySymmetricToroidal1988} have attracted strong interest, notably leading to the designs of the HSX \citep{andersonHelicallySymmetricExperiment1995} and NCSX \citep{zarnstorffPhysicsCompactAdvanced2001} stellarators. Multiple formulations of quasisymmetry exist \citep{helanderTheoryPlasmaConfinement2014, rodriguezNecessarySufficientConditions2020, burbyMathematicsQuasisymmetry2020}, all of which employ flux coordinates and therefore require the existence of nested flux surfaces. We propose a method of constructing approximate flux coordinates on an isolated flux surface, on which quasisymmetry can then be defined and optimised for. The existence of at least one isolated flux surface will be guaranteed, by imposing the boundary condition that the magnetic field be tangential on a prescribed boundary. Note that we will not consider whether this boundary condition can actually be realised with a set of coils, a task pursued by codes like FOCUS \citep{zhuNewMethodDesign2018}.

With the exception of \cite{landremanMagneticFieldsPrecise2021}, previous optimisation studies \citep{nuhrenbergQuasihelicallySymmetricToroidal1988, kuNewClassesQuasihelically2011, drevlakESTELLQuasiToroidallySymmetric2013, baderStellaratorEquilibriaReactor2019, hennebergPropertiesNewQuasiaxisymmetric2019, hennebergImprovingFastparticleConfinement2020, landremanStellaratorOptimizationGood2021} targeted quasisymmetry by minimising the symmetry-breaking components of the magnetic field strength in Boozer coordinates, often for vacuum magnetic fields. The most widely-used solver, whether for vacuum fields or plasmas with finite pressure, is the VMEC code \citep{hirshmanThreedimensionalFreeBoundary1986}, which notably assumes the existence of nested flux surfaces. 
We will employ the SPEC code \citep{hudsonComputationMultiregionRelaxed2012}, which does not build in this assumption. Furthermore, in contrast to most previous studies, we use a formulation of quasisymmetry that does not rely on a Boozer coordinate transformation, although it still enables the specification of a desired helicity of the magnetic field strength. 

Previous studies have sought to optimise for quasisymmetry either on a single flux surface \citep[e.g.][]{hennebergImprovingFastparticleConfinement2020}, or on multiple flux surfaces \citep[e.g.][]{landremanMagneticFieldsPrecise2021} with the aim of approximating quasisymmetry in a finite volume. We will herein consider quasisymmetry on a single flux surface only. Away from a surface with exact quasisymmetry, the deviation from quasisymmetry will generally increase linearly in the flux difference \citep{senguptaVacuumMagneticFields2021}. It will thus be of interest to extend the present work on vacuum fields to multi-region relaxed magnetohydrodynamic (MRxMHD) equilibria. In this model, the interfaces between force-free regions are flux surfaces, such that quasisymmetry can be optimised for on multiple flux surfaces.

This paper is structured as follows. We begin with a brief introduction to adjoint methods in \S\ref{sec:basics_adjoint_methods}. A method of constructing approximate flux coordinates on a single flux surface is introduced in \S\ref{sec:approximate_flux_coordinates}. The derived adjoint equations for vacuum fields are presented in \S\ref{sec:adjoint_formalism_vacuum_fields}, first for a simpler objective function targeting a given rotational transform value on the boundary in \S\ref{sec:iota_fom_shape_derivative}, then for one targeting quasisymmetry on the boundary with a given helicity in \S\ref{sec:QS_fom_shape_derivative}. The resulting shape gradients are evaluated numerically and benchmarked against finite-difference calculations.

%%%%%%%%%%%%%%%%%%%%%%%%%%%%%%%%%%%%%%%%%%%%%%%%%%%%%%%%%%%%%%%%%%%
\section{Basics of adjoint methods}
\label{sec:basics_adjoint_methods}

We are interested in obtaining derivative information for a functional $f(\mathcal{S}, u(\mathcal{S}))$, called hereafter the objective function. This functional depends on the surface $\mathcal{S}$ explicitly and also implicitly through the solution $u(\mathcal{S})$ to a partial differential equation (PDE) $\mathcal{P}(\mathcal{S},u) = 0$. Here, $\mathcal{P}$ is a general operator and $u$ is member of a Hilbert space with associated inner product $\langle \cdot , \cdot \rangle$, taken in our case to be the surface integral $\int_\mathcal{S} \diff S \;(\cdot)(\cdot)$. 

Consider a displacement of the surface $\mathcal{S}$ in the direction $\delta\mathbf{x}$ with magnitude $\epsilon$, resulting in a perturbed surface $\mathcal{S}_\epsilon = \{ \mathbf{x}_0 + \epsilon \delta\mathbf{x}(\mathbf{x}_0) : \mathbf{x}_0 \in \mathcal{S}\}$. The shape derivative of an arbitrary function $g(\mathcal{S})$ in the direction $\delta\mathbf{x}$ is now defined as
\begin{equation}
    \delta g(\mathcal{S})[\delta\mathbf{x}] = \lim_{\epsilon\rightarrow 0} \frac{g(\mathcal{S_\epsilon}) - g(\mathcal{S})}{\epsilon}.
\end{equation}
If $g$ depends only on the geometrical shape of the surface, the shape derivative $\delta g[\delta\mathbf{x}]$ will be a function of only the normal component $\normpert$ of the displacement, as any tangential component of $\delta\mathbf{x}$ leaves the shape of $\mathcal{S}$ unchanged to first order. Here, $\normvec$ is a normal vector on $\mathcal{S}$.

To compute derivatives of the objective function while enforcing the PDE constraint $\mathcal{P}(\mathcal{S},u) = 0$, the method of Lagrange multipliers is used. Consider the Lagrangian
\begin{equation}
    \mathcal{L}(\mathcal{S},u,q) = f(\mathcal{S},u) + \int_\mathcal{S} \diff S \; q \; \mathcal{P}(\mathcal{S},u),
\end{equation}
with the Lagrange multiplier $q$. Its shape derivative $\delta\mathcal{L}[\delta\mathbf{x}]$ contains explicit contributions in the perturbation $\delta\mathbf{x}$, as well as implicit contributions through $\delta q[\delta\mathbf{x}]$ and $\delta u[\delta\mathbf{x}]$. The implicit dependencies are removed by making $\mathcal{L}$ stationary with respect to $\delta q[\delta\mathbf{x}]$, which is equivalent to enforcing the original PDE $\mathcal{P}(\mathcal{S},u) = 0$, and $\delta u[\delta\mathbf{x}]$, which leads to an adjoint PDE for $q$. 

If $\mathcal{L}$ is stationary with respect to both $\delta u[\delta\mathbf{x}]$ and $\delta q[\delta\mathbf{x}]$, the remaining explicit dependence of its shape derivative $\delta\mathcal{L}(\mathcal{S},u,q)[\delta\mathbf{x}]$ is equal to the shape derivative of the figure of merit $\delta f(\mathcal{S},u(\mathcal{S}))[\delta\mathbf{x}]$ with $u=u(\mathcal{S})$ satisfying the PDE constraint, as shown in e.g. \citet{paulAdjointMethodsStellarator2020}. The Hadamard-Zol\'esio structure theorem \citep{delfourShapesGeometries2011} further states that the remaining contribution to the Lagrangian's shape derivative, provided $\mathcal{L}$ is sufficiently smooth, can be expressed as
\begin{equation}
    \delta\mathcal{L}(\mathcal{S},u,q)[\delta\mathbf{x}] = \int_\mathcal{S} \diff S \;(\normpert)\; \mathcal{G},
\end{equation}
where $\mathcal{G}$ is called the shape gradient, and can be interpreted as the local sensitivity of the objective function to perturbations of $\mathcal{S}$.

In practice, the surface $\mathcal{S}$ is typically represented by a finite set of parameters $\Omega = \{\Omega_i,\; i = 1, 2, \dots N\}$, e.g. Fourier coefficients $\{R_{m,n}, Z_{m,n}\}$, and the functional $f(\mathcal{S},u(\mathcal{S}))$ is approximated by a function $f(\Omega,u(\Omega))$. The derivative of $f(\Omega,u(\Omega))$ with respect to a parameter $\Omega_i$ can be approximated as
\begin{equation}
    \frac{\partial f(\Omega, u(\Omega))}{\partial \Omega_i} = \int_\mathcal{S} \diff S\; \frac{\partial \mathbf{x}}{\partial \Omega_i}\cdot \normvec \; \mathcal{G}. \label{eq:parameter_derivatives_shape_grad}
\end{equation}
The adjoint method of evaluating the parameter derivatives required for optimisation or sensitivity analysis \citep{paulAdjointMethodsStellarator2020} thus consists in computing $\delta\mathcal{L}[\delta\mathbf{x}]$ to obtain the adjoint PDE for $q$ and the shape gradient $\mathcal{G}$, which is then used to evaluate the right-hand-side of \eqref{eq:parameter_derivatives_shape_grad}.

When evaluating the parameter derivatives numerically through \eqref{eq:parameter_derivatives_shape_grad}, errors are introduced from the inexact solutions to the original and adjoint PDEs. Indeed, these PDEs are assumed to be exactly satisfied in the preceding derivation to remove the implicit dependencies of $\delta\mathcal{L}[\delta\mathbf{x}]$ on $\delta u[\delta\mathbf{x}]$ and $\delta q[\delta\mathbf{x}]$, and also typically when deriving an expression for the shape gradient $\mathcal{G}$.

The formalism presented above can easily be generalised to multiple PDE constraints, and, for a closed $\mathcal{S}$, to PDEs satisfied not on $\mathcal{S}$ but in the volume enclosed by it. This will be done in \S\ref{sec:adjoint_formalism_vacuum_fields}, where both the Laplace equation for the vacuum field and the straight field line equation, respectively valid in the volume and on the boundary, will be enforced as constraints, with two corresponding adjoint variables.

%%%%%%%%%%%%%%%%%%%%%%%%%%%%%%%%%%%%%%%%%%%%%%%%%%%%%%%%%%%%%%%%%%%
\section{Evaluating approximate flux coordinates on an isolated flux surface}
\label{sec:approximate_flux_coordinates}

The existence of nested flux surfaces is commonly assumed in theoretical studies of magnetically confined plasmas, e.g. to formulate quasisymmetry. In particular, many formulas involve $\nabla\psi$, where the toroidal flux $\psi$ is a global flux surface label. However, three-dimensional magnetic fields lacking a continuous symmetry are not generally integrable. It is desirable to generalise $\nabla\psi$ to the case of an isolated flux surface, i.e. a flux surface in whose neighbourhood the field is generally non-integrable.

On a flux surface $\mathcal{S}$, the magnetic field's normal component vanishes by definition, i.e.  $\mathbf{B}\cdot\normvec = 0$ with $\normvec$ the unit normal vector on $\mathcal{S}$. The field line label $\alpha$ on $\mathcal{S}$ is defined through the straight field line equation $\mathbf{B}\cdot\nabla_\Upgamma\alpha = 0$. Here, the tangential gradient $\nabla_\Upgamma$, defined in App.~\ref{app:diff_operators_surfaces}, is the component of the 3D gradient tangential to the surface \eqref{eq:def_tangential_gradient}. Note that in the integrable case, $\nabla\psi$ is normal to the flux surfaces, and the magnetic field satisfies $\mathbf{B} = \nabla\psi\times\nabla\alpha$.

We now define the generalisation $\overline{\nabla\psi}$ on $\mathcal{S}$ of the toroidal flux gradient $\nabla\psi$, by setting $\overline{\nabla\psi}$ normal to $\mathcal{S}$, and by requiring $\mathbf{B} = \overline{\nabla\psi}\times\nabla\alpha$ to be satisfied on $\mathcal{S}$. Squaring the latter equality and using $\overline{\nabla\psi} = \normvec |\overline{\nabla\psi}|$ yields
\begin{equation}
    \overline{\nabla\psi} = \normvec \; \frac{B}{\abs{\nabla_\Upgamma \alpha}}, \label{eq:def_nabla_tilde_psi}
\end{equation}
where $B = \abs{\mathbf{B}}$ is the magnetic field strength. Note that $\overline{\nabla\psi}$ is defined through \eqref{eq:def_nabla_tilde_psi}, and should not be misinterpreted as the gradient of a scalar function.

The defining expression for $\overline{\nabla\psi}$ \eqref{eq:def_nabla_tilde_psi} can be evaluated on any flux surface without requiring nested flux surfaces in its neighbourhood, and will revert to $\overline{\nabla\psi} = \nabla\psi$ when the field is integrable in the neighbourhood of that flux surface. In practice, one might couple objective functions relying on \eqref{eq:def_nabla_tilde_psi} with a figure of merit targeting integrability, aiming for a final plasma shape for which the field is integrable, such that $\overline{\nabla\psi} = \nabla\psi$ and the minimised objective function represents the physical quantity of interest.

\begin{figure}

\begin{subfigure}{.5\textwidth}
  \centering
  \includegraphics[width=1\linewidth]{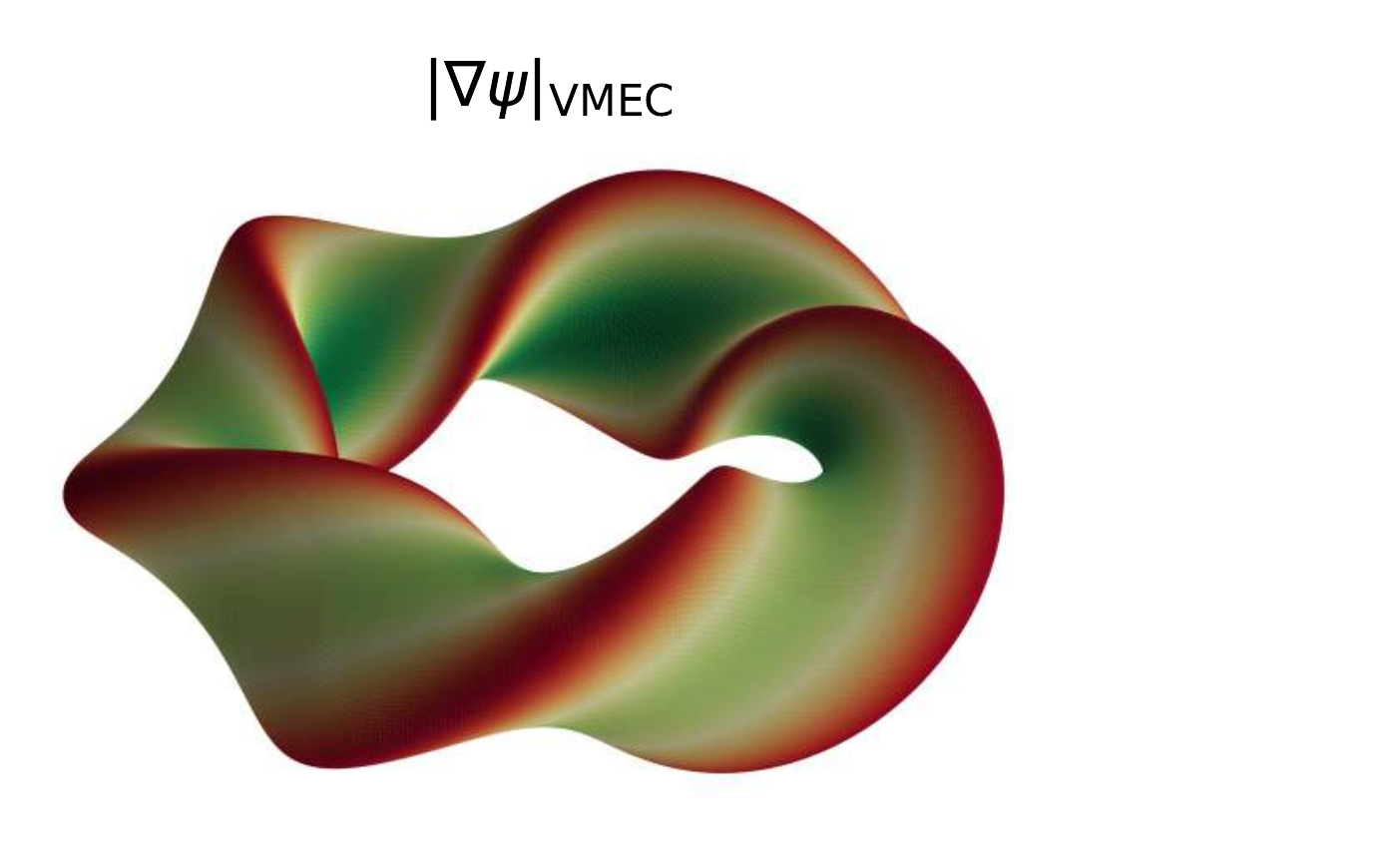}
  \caption{}
  \label{fig:abs_nabla_psi_VMEC}
\end{subfigure}%
\begin{subfigure}{.5\textwidth}
  \centering
  \includegraphics[width=1\linewidth]{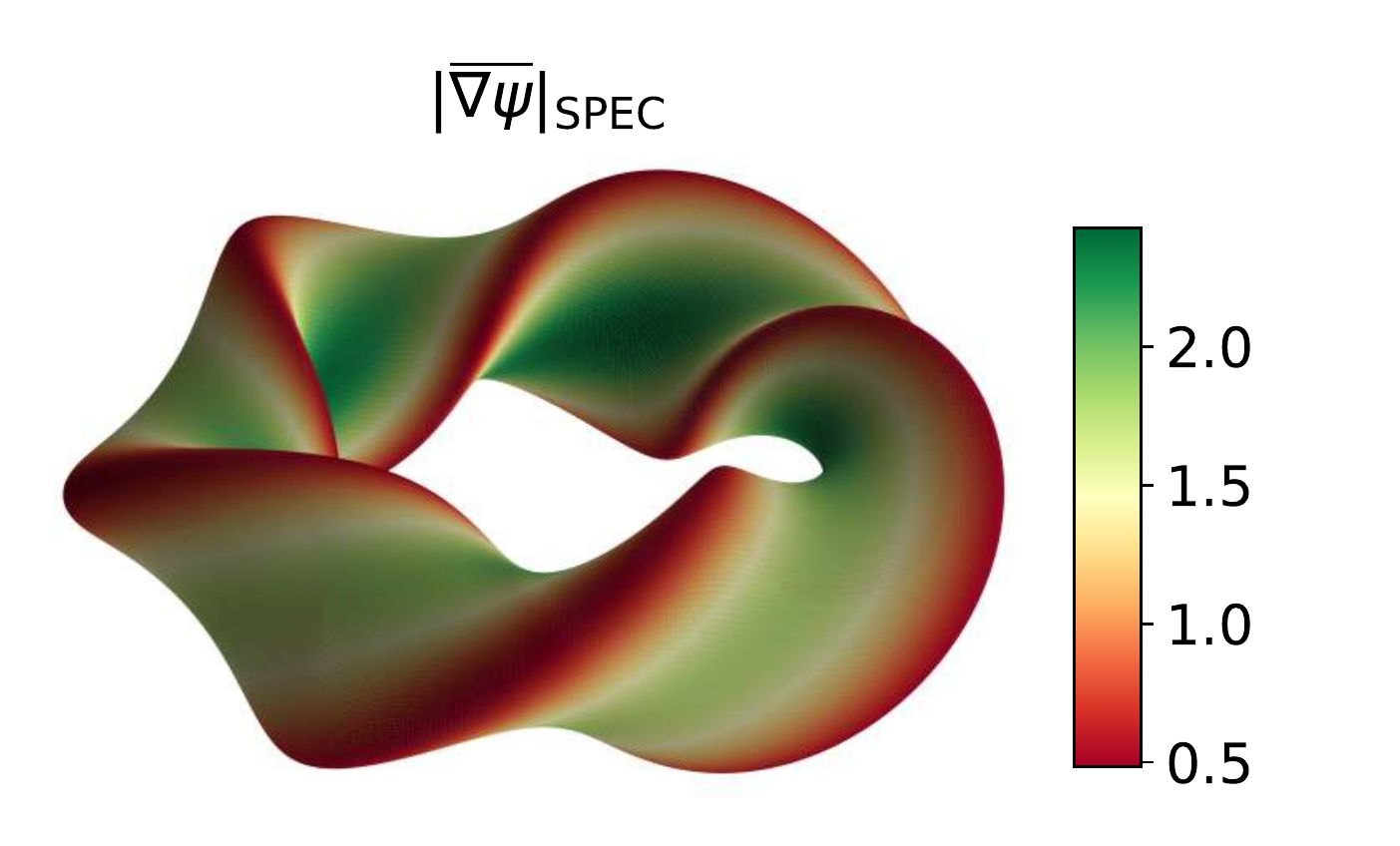}
  \caption{}
  \label{fig:abs_nabla_psi_SPEC}
\end{subfigure}\\
\begin{subfigure}{.5\textwidth}
  \centering
  \includegraphics[width=1\linewidth]{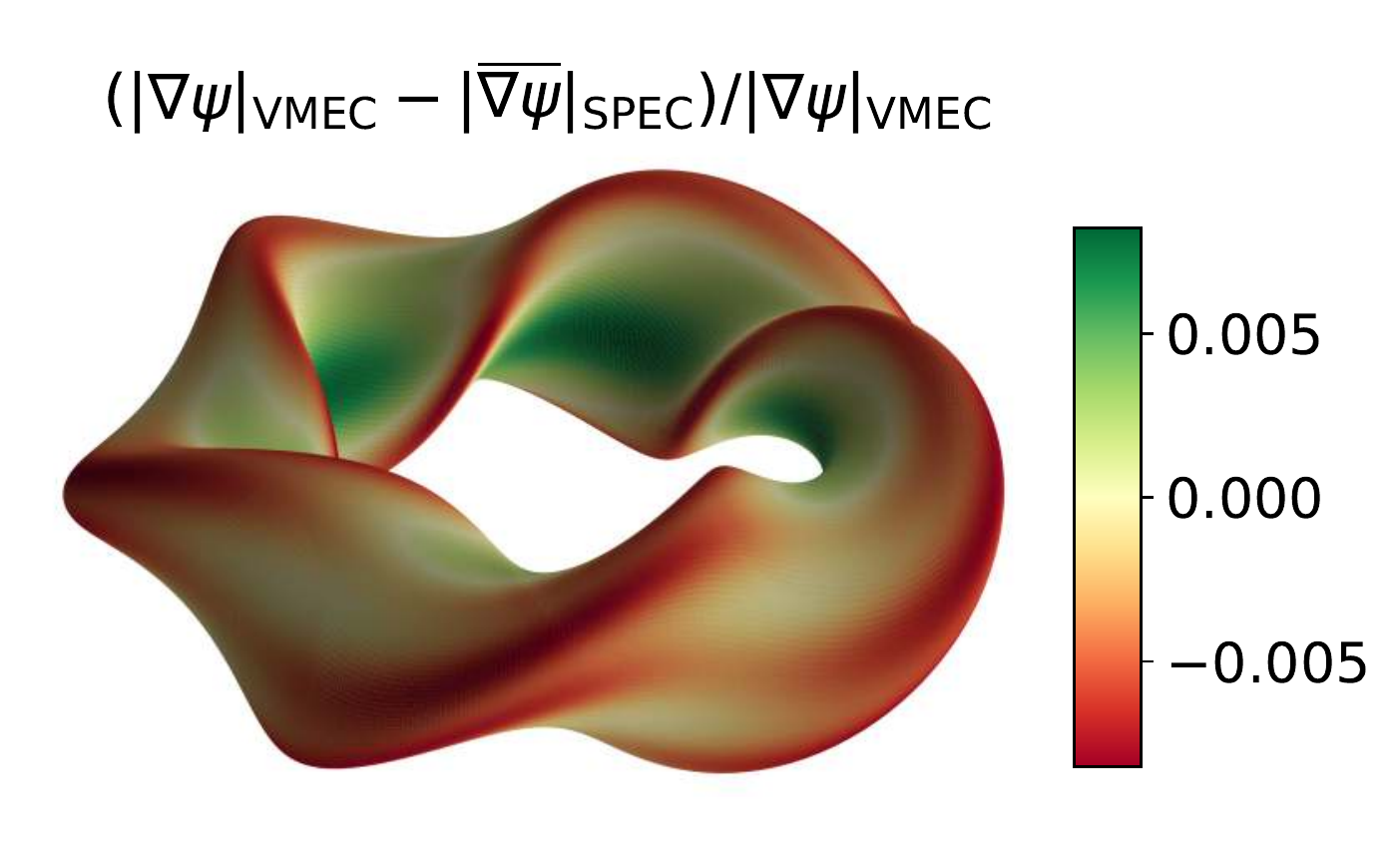}
  \caption{}
  \label{fig:relative_diff_abs_nabla_psi}
\end{subfigure}%
\begin{subfigure}{.5\textwidth}
  \centering
  \includegraphics[width=1\linewidth]{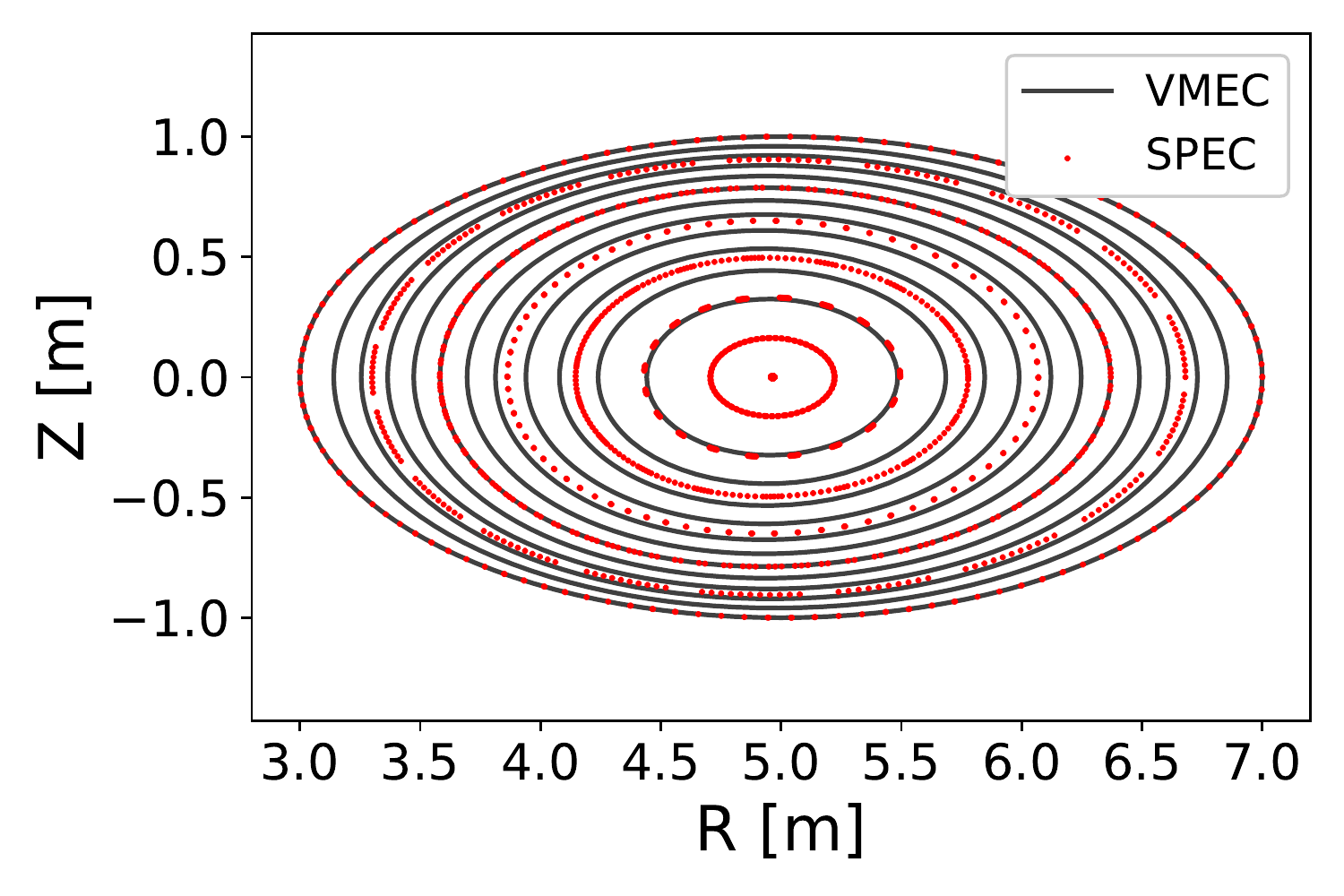}
  \caption{}
  \label{fig:poincare_surfaces_VMEC_SPEC}
\end{subfigure}%
\caption{Comparison of (a) the toroidal flux gradient $\abs{\nabla\psi}$ evaluated with VMEC with (b) the generalised toroidal flux gradient $|\overline{\nabla\psi}|$ \eqref{eq:def_nabla_tilde_psi} obtained in SPEC, for a $5$-period rotating ellipse case with half a rotation per field period, major radius at the ellipse centre $R_0 = 5$\;m, and ellipse major and minor axes values of $2$\;m and $1$\;m, respectively. The relative difference between the two quantities is below $1\%$, as shown in (c). A small difference is to be expected in this case, where integrability is well satisfied, as attested in (d) by the Poincar\'e plot at toroidal angle $\phi=0$ from the SPEC calculation, which agrees well with the flux surfaces computed by VMEC. All data generated in this paper can be obtained from \cite{niesDataCodesPaper2021}.}
\label{fig:abs_nabla_psi}
\end{figure}

The generalised toroidal flux gradient \eqref{eq:def_nabla_tilde_psi} is evaluated in Fig.~\ref{fig:abs_nabla_psi_SPEC} for a rotating ellipse configuration computed with the SPEC code. It agrees excellently with the toroidal flux gradient evaluated by VMEC, which can be calculated directly due to the imposed nestedness of flux surfaces, shown in Fig.~\ref{fig:abs_nabla_psi_VMEC}. The relative difference between the two is below a percentage point in this case, as shown in Fig.~\ref{fig:relative_diff_abs_nabla_psi}. The difference is expected to be small when integrability is satisfied, which indeed seems to hold here, as attested by the absence of islands and chaotic regions in the SPEC Poincar\'e plot shown in Fig.~\ref{fig:poincare_surfaces_VMEC_SPEC}. Note that SPEC solves for a vacuum magnetic field, while VMEC computes an ideal MHS equilibrium with vanishing thermal pressure, and with a plasma current that is small but finite due to the constraint of integrability.

The generalised toroidal flux gradient $\overline{\nabla\psi}$ can be applied generally in any situation where local flux coordinates need to be evaluated, e.g. in calculations of perpendicular transport or magnetohydrodynamic stability. Isolated flux surfaces occur e.g. in fixed-boundary equilibrium calculations, where the plasma outer boundary is constrained to be a flux surface as a boundary condition on the magnetic field, or at the interfaces of MRxMHD equilibria computed by e.g. SPEC \citep{hudsonComputationMultiregionRelaxed2012} or BIEST \citep{malhotraTaylorStatesStellarators2019}. In the following, we will employ \eqref{eq:def_nabla_tilde_psi} specifically for a fixed-boundary vacuum field to formulate quasisymmetry on the boundary.

%%%%%%%%%%%%%%%%%%%%%%%%%%%%%%%%%%%%%%%%%%%%%%%%%%%%%%%%%%%%%%%%%%%
\section{Application of adjoint formalism to vacuum fields}
\label{sec:adjoint_formalism_vacuum_fields}

Consider a vacuum magnetic field $\mathbf{B}$ in a toroidal domain $\mathcal{V}$ bounded by the surface $\mathcal{S} = \partial \mathcal{V}$. As the vacuum magnetic field is curl-free, it can be expressed as $\mathbf{B}=\nabla\Phi$, with the scalar potential $\Phi$. Because we consider a simple torus $\mathcal{V}$, the most general form for the scalar potential is $\Phi=G (\omega+\phi)$, where $G$ is a constant, $\omega$ is a single-valued function on $\mathcal{S}$, and $\phi$ is an arbitrary toroidal angle. By integrating the magnetic field along a toroidal loop around the torus, the constant $G$ is found to be proportional to the net external current through the `hole' of the torus.

As the magnetic field is divergence-less, the magnetic scalar potential satisfies the Laplace equation. The field's normal component is constrained to vanish on $\mathcal{S}$ by imposing a Neumann boundary condition on the magnetic scalar potential. Further prescribing e.g. $G$, or the toroidal flux, guarantees a unique solution to Laplace's equation. We herein opt to hold the toroidal flux fixed, although the shape derivative $\delta G[\delta\mathbf{x}]$ will not appear in this study due to our normalisation of the figure of merit for quasisymmetry \eqref{eq:definition_fQS}. A different choice of normalisation would lead to an additional contribution proportional to $\delta G[\delta\mathbf{x}]$ in the shape derivative of the Lagrangian.

For convenience, we define the normalised magnetic field $\mathbf{\breve{B}}$ as
\begin{equation}
    \mathbf{\breve{B}} \equiv \mathbf{B}/G = \nabla\big(\omega + \phi\big). \label{eq:magnetic_field_scalar_pot}
\end{equation}
Let us further assume the toroidal angle $\phi$ to be the azimuthal angle in cylindrical coordinates, satisfying $\Updelta\phi=0$ in the domain of interest. We can thus write
\begin{subequations}
\begin{align}
    \nabla\cdot\mathbf{\breve{B}} = \Updelta\omega = 0   \qquad\qquad\qquad\mathrm{in}\;&\mathcal{V}, \label{eq:vacuum_field_Laplace}\\
    \mathbf{\breve{B}}\cdot\normvec = \nabla(\omega+\phi)\cdot\normvec = 0 \qquad\mathrm{on}\; & \mathcal{S} \label{eq:vacuum_field_normal_BC},
\end{align}
\end{subequations}
with $\normvec$ the normal unit vector on $\mathcal{S}$. Furthermore, the shape derivative $\delta\omega[\delta\mathbf{x}]$ satisfies
\begin{subequations}
\begin{align}
    \Updelta(\delta\omega[\delta\mathbf{x}]) = 0   \qquad\qquad\mathrm{in}\;&\mathcal{V}, \label{eq:vacuum_field_pert_Laplace}\\
    \mathbf{\breve{B}}\cdot\delta\normvec[\delta\mathbf{x}] + \nabla(\delta\omega[\delta\mathbf{x}])\cdot\normvec + (\normpert)\;\normvec\cdot\nabla \mathbf{\breve{B}}\cdot\normvec = 0 \quad\mathrm{on}\; & \mathcal{S} \label{eq:vacuum_field_pert_normal_BC},
\end{align}
\end{subequations}
where the Laplace equation is obtained from noting the commutative property of shape and spatial derivatives, and the normal boundary condition on $\delta\omega$ was derived in e.g. \citet[\S 3.2]{sokolowskiIntroductionShapeOptimization1992}. The shape derivative of the normal vector $\delta\normvec[\delta\mathbf{x}]=-\nabla_\Upgamma(\normpert)$ is derived in App.~\ref{app:normal_extension_and_normal_vector_shape_derivative}.

Evaluating the rotational transform and quasisymmetry figures of merit further requires the solution to the straight field line equation
\begin{equation}
    0 = \mathbf{B}\cdot\nabla_\Upgamma\alpha \qquad\mathrm{on}\; \mathcal{S}, \label{eq:straight_field_line_eq}
\end{equation}
with the field line label
\begin{equation}
   \alpha \equiv \theta-\iota\phi+\lambda(\theta,\phi), \label{eq:def_field_line_label_alpha}
\end{equation}
where $\theta$ is a general poloidal angle, $\lambda$ is a single-valued function of $\theta$ and $\phi$, and $\iota$ is a scalar. Note that both $\lambda$ and $\iota$ are defined on the boundary $\mathcal{S}$ only, through \eqref{eq:def_field_line_label_alpha}.

Let us define the Lagrangian corresponding to an arbitrary objective function $f(\mathcal{S},\omega,\iota,\lambda)$,
\begin{equation}
    \mathcal{L}(\mathcal{S}, \omega, q_\omega, \iota, \lambda, q_\alpha ) = f(\mathcal{S},\omega,\iota,\lambda) + \mathcal{M}(\mathcal{S}, \omega, q_\omega) + \mathcal{N}(\mathcal{S}, \omega, \iota, \lambda, q_\alpha), \label{eq:lagrangian_general}
\end{equation}
with the weak form of the Laplace equation \eqref{eq:vacuum_field_Laplace}
\begin{equation}
    \mathcal{M}(\mathcal{S}, \omega, q_\omega) = \int_{\mathcal{V}} \diff V \; q_\omega \Updelta \omega, \label{eq:weak_form_Laplace_vac}
\end{equation}
and the weak form of the straight field line equation \eqref{eq:straight_field_line_eq} normalised by $G$ 
\begin{equation}
    \mathcal{N}(\mathcal{S}, \omega, \iota, \lambda, q_\alpha) = \int_{\mathcal{S}} \diff S \; q_\alpha \mathbf{\breve{B}} \cdot \nabla_\Upgamma \alpha. \label{eq:weak_form_straight_field_line_eq}
\end{equation}
As explained in \S\ref{sec:basics_adjoint_methods}, $q_\omega$ and $q_\alpha$ act as Lagrange multipliers: making the Lagrangian \eqref{eq:lagrangian_general} stationary with respect to $\delta q_\omega[\delta\mathbf{x}]$ and $\delta q_\alpha[\delta\mathbf{x}]$ ensures that the Laplace \eqref{eq:vacuum_field_Laplace} and straight field line \eqref{eq:straight_field_line_eq} equations are satisfied, respectively. These trivial variations are omitted in the following under the assumption that \eqref{eq:vacuum_field_Laplace} and \eqref{eq:straight_field_line_eq} are satisfied, thus considering only the implicit dependencies of $\delta\mathcal{L}[\delta\mathbf{x}]$ on $\delta\omega[\delta\mathbf{x}]$, $\delta\iota[\delta\mathbf{x}]$, and $\delta\lambda[\delta\mathbf{x}]$ to obtain the adjoint equations for $q_\omega$ and $q_\alpha$.

First, the shape derivative of $\mathcal{M}$ \eqref{eq:weak_form_Laplace_vac}, derived in App.~\ref{app:derivation_laplace_eq_shape_derivative}, is
\begin{equation}
    \delta\mathcal{M}[\delta\mathbf{x}] = \int_{\mathcal{V}} \diff V \; \delta\omega[\delta\mathbf{x}] \Updelta q_\omega - \int_{\mathcal{S}} \diff S \; \bigg[ \delta \omega[\delta\mathbf{x}]  \nabla q_\omega \cdot \normvec - (\normpert)\;  \mathbf{\breve{B}}\cdot\nabla q_\omega  \bigg]. \label{eq:variation_M_tot}
\end{equation}

Second, the shape derivative of $\mathcal{N}$ \eqref{eq:weak_form_straight_field_line_eq}, derived in App.~\ref{app:derivation_straight_field_line_eq_shape_derivative}, is
\begin{align}
    \delta\mathcal{N}[\delta\mathbf{x}]  = \int_{\mathcal{S}} & \diff S\;  \bigg[ - \delta\omega[\delta\mathbf{x}] \;\nabla_\Upgamma \cdot \left(q_\alpha \nabla_\Upgamma \alpha\right) - \delta\iota[\delta\mathbf{x}]\; q_\alpha \mathbf{\breve{B}} \cdot \nabla\phi   \label{eq:variation_N_tot}\\
    \nonumber & - \delta\lambda[\delta\mathbf{x}]\; \nabla_\Upgamma \cdot \left( q_\alpha \mathbf{\breve{B}} \right) + (\normpert) q_\alpha \left( \normvec\cdot\nabla\mathbf{\breve{B}}\cdot\nabla_\Upgamma\alpha - \mathbf{\breve{B}}\cdot\nabla\normvec\cdot\nabla_\Upgamma\alpha \right) \bigg].
\end{align}
The tangential gradient $\nabla_\Upgamma(\cdot)$ and tangential divergence $\nabla_\Upgamma\cdot(\cdot)$ operators are defined in App.~\ref{app:diff_operators_surfaces}.

We now proceed by computing the shape derivatives of two objective functions, first targeting a given rotational transform value on $\mathcal{S}$ (\S\ref{sec:iota_fom_shape_derivative}), and second targeting quasi-symmetry on $\mathcal{S}$ with a given helicity value (\S\ref{sec:QS_fom_shape_derivative}). We will then be able to evaluate the shape derivative of the Lagrangian \eqref{eq:lagrangian_general}, yielding the adjoint equations and shape gradient formulas. Numerical verification and example shape gradients are shown for each figure of merit.

\subsection{Rotational transform objective function}
\label{sec:iota_fom_shape_derivative}

Before evaluating the more complicated shape gradient for the quasisymmetry figure of merit in \S\ref{sec:QS_fom_shape_derivative}, we consider a simple figure of merit targeting a given target rotational transform $\iota_T$ on the surface $\mathcal{S}$. We thus define
\begin{equation}
    f_\iota(\iota) = \frac{1}{2} ( \iota - \iota_T )^2  \label{eq:f_iota},
\end{equation}
where $\iota$ is the rotational transform on $\mathcal{S}$, obtained by solving the straight field line equation \eqref{eq:straight_field_line_eq}. The shape derivative of $f_\iota$ is simply
\begin{equation}
    \delta f_\iota [\delta\mathbf{x}] = \delta \iota[\delta\mathbf{x}] \; ( \iota - \iota_T ). \label{eq:variation_f_iota_tot}
\end{equation}
By combining \eqref{eq:variation_M_tot}, \eqref{eq:variation_N_tot}, and \eqref{eq:variation_f_iota_tot}, we obtain the shape derivative of the Lagrangian $\mathcal{L}_\iota$ [\eqref{eq:lagrangian_general} with $f=f_\iota$],
\begin{align}
    \delta\mathcal{L}&_\iota[\delta\mathbf{x}]  = \int_{\mathcal{V}} \diff V \; \delta\omega[\delta\mathbf{x}] \; \Updelta q_\omega - \int_{\mathcal{S}} \diff S \;  \delta \omega [\delta\mathbf{x}] \bigg[ \nabla q_\omega \cdot \normvec + \nabla_\Upgamma \cdot \Big(q_\alpha \nabla_\Upgamma \alpha \Big) \bigg] \label{eq:variation_L_iota_tot}\\
    \nonumber & - \delta\iota[\delta\mathbf{x}]\; \bigg[ \int_{\mathcal{S}} \diff S \; q_\alpha \mathbf{\breve{B}} \cdot \nabla\phi + ( \iota - \iota_T ) \bigg] - \int_{\mathcal{S}} \diff S \;\delta\lambda[\delta\mathbf{x}]\; \nabla_\Upgamma \cdot \Big( q_\alpha \mathbf{\breve{B}} \Big)  \\
    \nonumber & + \int_{\mathcal{S}} \diff S \;(\normpert) \bigg[\nabla q_\omega \cdot \mathbf{\breve{B}} + q_\alpha \left( \normvec\cdot\nabla\mathbf{\breve{B}}\cdot\nabla_\Upgamma\alpha - \mathbf{\breve{B}}\cdot\nabla\normvec\cdot\nabla_\Upgamma\alpha \right) \bigg] \Bigg].
\end{align}

First, we obtain the adjoint equation for $q_\alpha$ by requiring the second line of \eqref{eq:variation_L_iota_tot} to vanish,
\begin{subequations}
\begin{align}
   \nabla_\Upgamma \cdot \Big( q_\alpha \mathbf{\breve{B}} \Big) = 0 \label{eq:iota_fom_adjoint_diff_eq_qsfl}, \\
   \int_{\mathcal{S}} \diff S \; q_\alpha \mathbf{\breve{B}} \cdot \nabla\phi + ( \iota - \iota_T ) = 0 \label{eq:iota_fom_adjoint_integral_eq_qsfl},
\end{align}
\end{subequations}
with both equations defined on $\mathcal{S}$. Using \eqref{eq:surface_divergence_theorem}, the surface integral of \eqref{eq:iota_fom_adjoint_diff_eq_qsfl} yields $0 = \mathbf{\breve{B}}\cdot\normvec$, which is consistent with the boundary condition on the magnetic field \eqref{eq:vacuum_field_normal_BC}. The first equation \eqref{eq:iota_fom_adjoint_diff_eq_qsfl} can be recast in the form of a magnetic differential equation $\mathbf{B}\cdot\nabla q_\alpha = - q_\alpha\nabla_\Upgamma\cdot\mathbf{B}$, while the second equation \eqref{eq:iota_fom_adjoint_integral_eq_qsfl} is an integral condition on $q_\alpha$ that ensures uniqueness of the solution.

Second, the adjoint equation for $q_\omega$ is obtained by requiring the first line of \eqref{eq:variation_L_iota_tot} to vanish,
\begin{subequations}
\begin{align}
    \Updelta q_\omega = 0   \qquad\qquad\qquad\mathrm{in}\;&\mathcal{V}, \label{eq:iota_fom_adjoint_eq_qomega}\\
    \nabla q_\omega\cdot\normvec = -\nabla_\Upgamma \cdot \Big(q_\alpha \nabla_\Upgamma \alpha \Big) \qquad\mathrm{on}\; & \mathcal{S} \label{eq:iota_fom_adjoint_eq_qomega_normal_BC}.
\end{align}
\end{subequations}
Like the magnetic potential $\omega$, the adjoint variable $q_\omega$ satisfies the Laplace equation in $\mathcal{V}$ \eqref{eq:iota_fom_adjoint_eq_qomega}. However, contrary to $\omega$, $q_\omega$ has a non-zero normal boundary condition on $\mathcal{S}$ \eqref{eq:iota_fom_adjoint_eq_qomega_normal_BC}, which notably depends on the straight field line adjoint variable $q_\alpha$. Equations \eqref{eq:iota_fom_adjoint_eq_qomega_normal_BC} and \eqref{eq:iota_fom_adjoint_eq_qomega} are consistent, as $\int_\mathcal{V} \diff V\; \Updelta q_\omega = \int_\mathcal{S} \diff S\; \nabla q_\omega\cdot\normvec = 0$, by \eqref{eq:surface_divergence_theorem}.

Finally, the remaining contribution from the last line of \eqref{eq:variation_L_iota_tot} yields the shape gradient
\begin{equation}
   \mathcal{G_\iota} = \frac{1}{G} \Big[ \mathbf{B}\cdot \nabla q_\omega + q_\alpha \big( \normvec\cdot\nabla\mathbf{B} - \mathbf{B}\cdot\nabla\normvec \big)\cdot\nabla_\Upgamma \alpha \Big], \label{eq:shape_gradient_iota_fom}
\end{equation}
with $\delta\mathcal{L}_\iota[\delta\mathbf{x}]  = \int_{\mathcal{S}} \diff S \;(\normpert)\;\mathcal{G_\iota}$.

We now calculate the shape gradient numerically and verify it against a finite-difference evaluation. The solutions to Laplace's equation for the vacuum magnetic field \eqref{eq:vacuum_field_Laplace} and adjoint equation for $q_\omega$ \eqref{eq:iota_fom_adjoint_eq_qomega} are calculated with the SPEC code \citep{hudsonComputationMultiregionRelaxed2012}, employing the new Zernike polynomial implementation \citep{quCoordinateParameterisationSpectral2020}. In all results shown, the radial resolution $L_\mathrm{rad}$ in SPEC is tied to the poloidal Fourier resolution $M_\mathrm{pol}$ through $L_\mathrm{rad} = M_\mathrm{pol} + 4$. The solutions to the straight field line and $q_\alpha$ adjoint equations are obtained with a Fourier-Galerkin spectral solver.

\begin{figure}
\begin{subfigure}{.5\textwidth}
  \centering
  \includegraphics[width=1\linewidth]{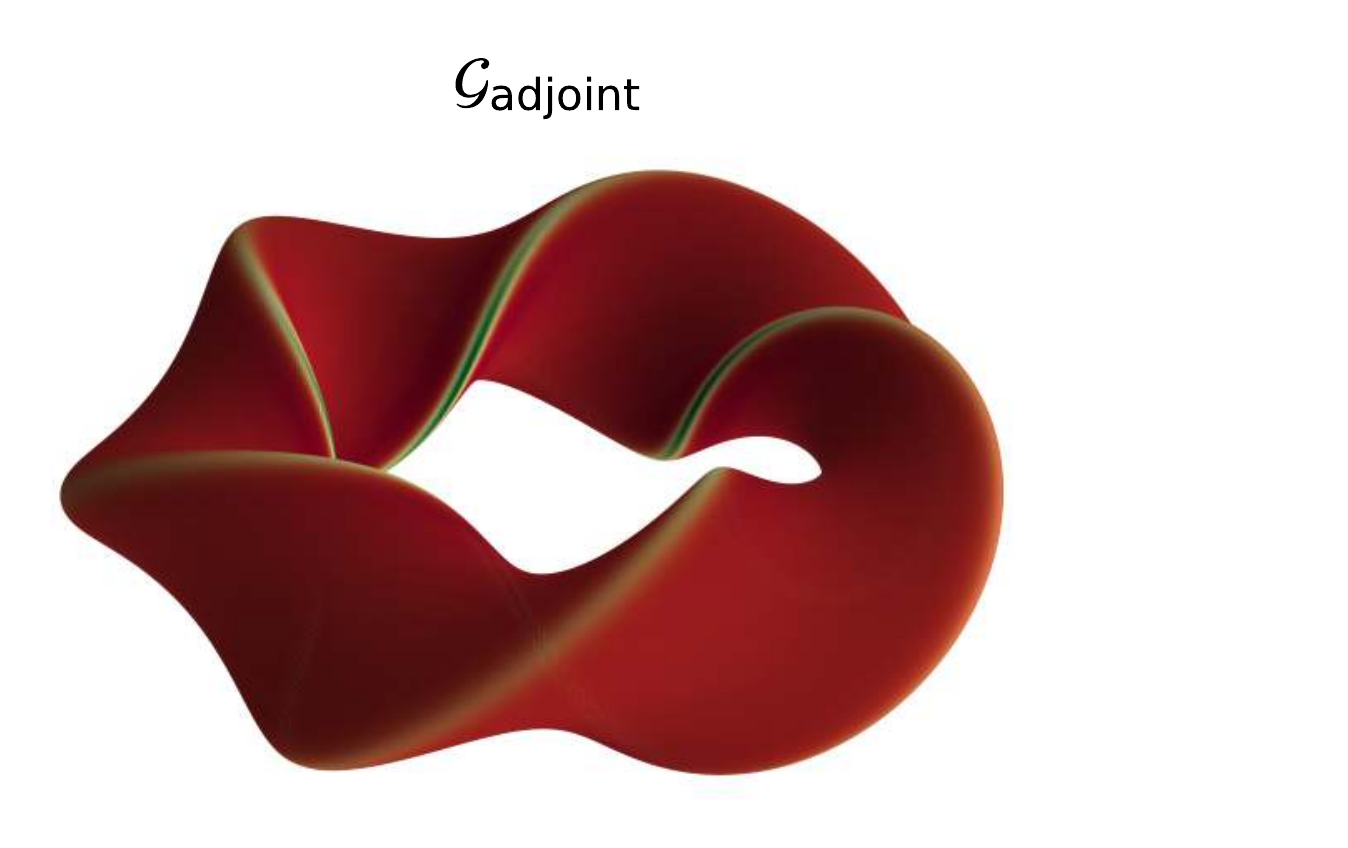}
  \caption{}
  \label{fig:shape_grad_iota_fom_adjoint}
\end{subfigure}%
\begin{subfigure}{.5\textwidth}
  \centering
  \includegraphics[width=1\linewidth]{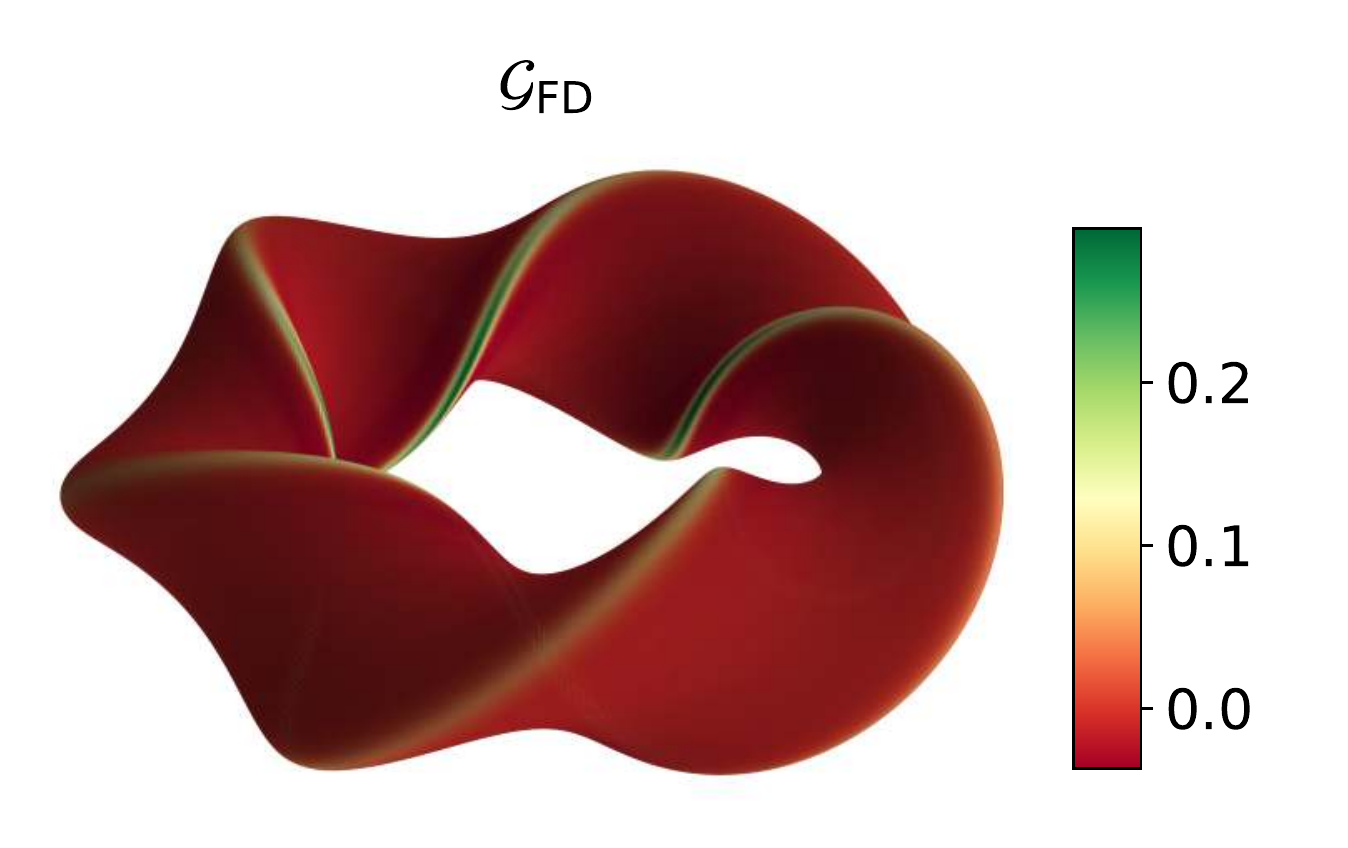}
  \caption{}
  \label{fig:shape_grad_iota_fom_findiff}
\end{subfigure}
\\
\begin{subfigure}{.5\textwidth}
  \centering
  \includegraphics[width=1\linewidth]{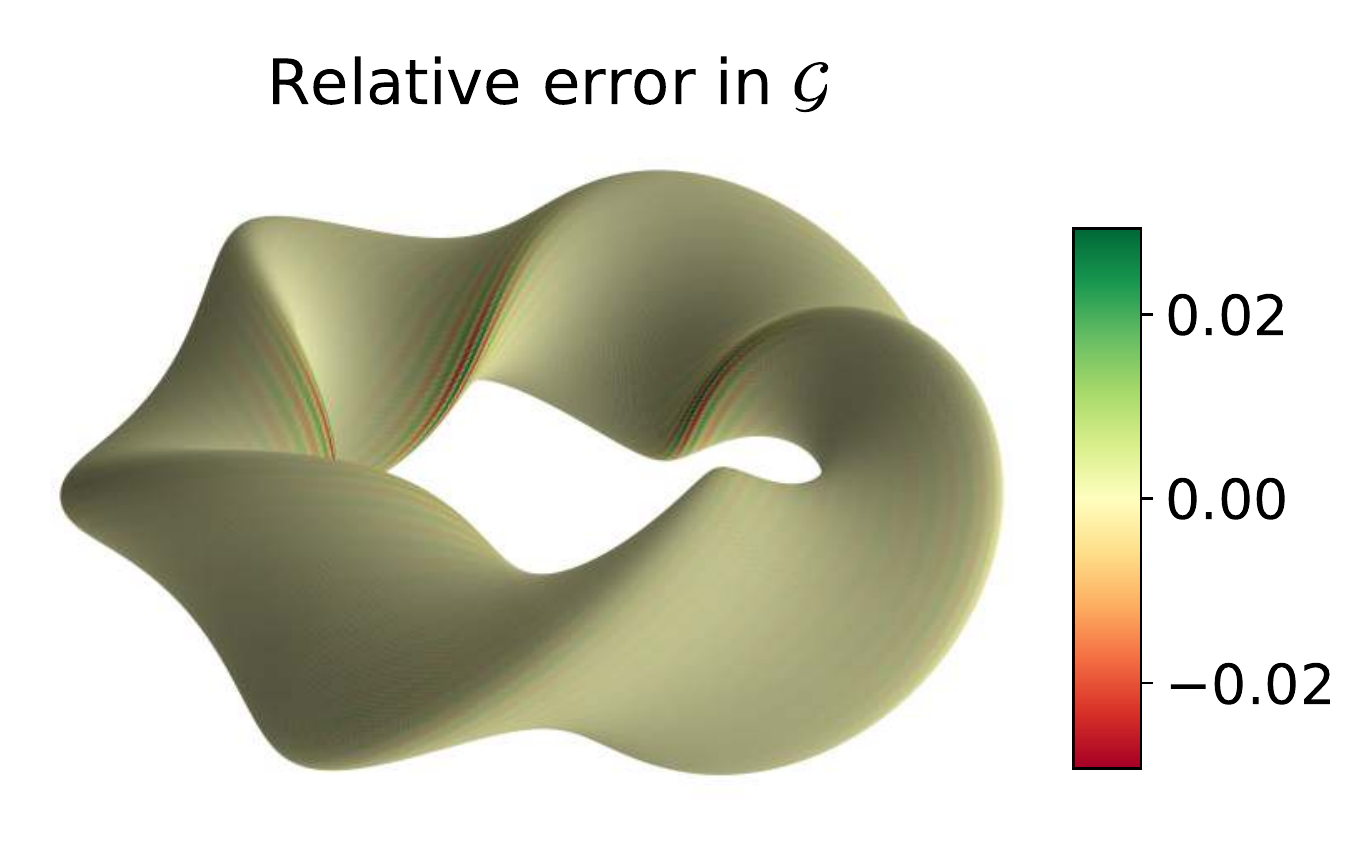}
  \caption{}
  \label{fig:shape_grad_iota_fom_relative_error}
\end{subfigure}%
\begin{subfigure}{.5\textwidth}
  \centering
  \includegraphics[width=1\linewidth]{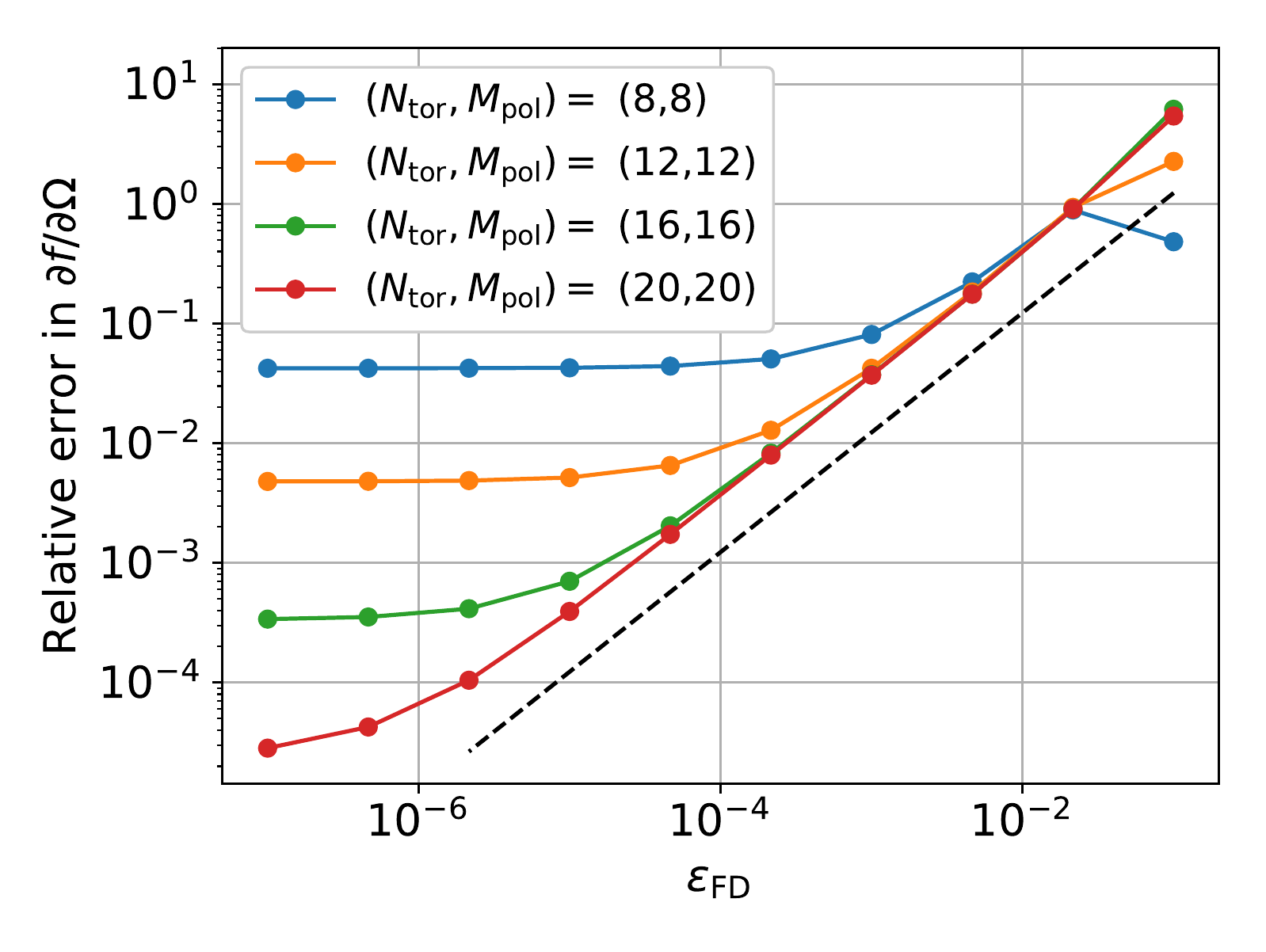}
  \caption{}
  \label{fig:convergence_iota_fom}
\end{subfigure}
\caption{Shape gradient for the rotational transform objective function with $\iota_T = 1$, evaluated through (a) adjoint methods and (b) a forward finite-difference scheme with step size $\epsilon_\mathrm{FD}=10^{-7}$, for the example rotating ellipse case introduced in Fig.~\ref{fig:abs_nabla_psi}, with Fourier resolution $(N_\mathrm{tor}, M_\mathrm{pol}) = (16,16)$. The relative error, defined as the absolute error normalised by the maximal absolute value of the adjoint shape gradient, is shown in (c). The convergence of the relative error in the parameter derivative \eqref{eq:parameter_derivatives_shape_grad} for a random direction in $\Omega$ is shown in (d) as a function of the step-size $\epsilon_\mathrm{FD}$ and Fourier resolution $(N_\mathrm{tor}, M_\mathrm{pol})$. The black dashed line indicates the linear scaling in $\epsilon_\mathrm{FD}$ expected from the employed forward finite-difference scheme.}
\label{fig:iota_fom_numerical_eval_and_convergence}
\end{figure}

The shape gradient \eqref{eq:shape_gradient_iota_fom} is shown in Fig.~\ref{fig:shape_grad_iota_fom_adjoint} for the example rotating ellipse case introduced in Fig.~\ref{fig:abs_nabla_psi}. The localisation at the ellipse tips is unsurprising, as near-axis expansions show that ellipticity of the flux surfaces generates rotational transform \citep{mercierEquilibriumStabilityToroidal1964}. This shape gradient $\mathcal{G}_\mathrm{adjoint}$ can be verified against the direct finite-difference evaluation $\mathcal{G}_\mathrm{FD}$ shown in Fig.~\ref{fig:shape_grad_iota_fom_findiff}, obtained by evaluating the parameter derivatives $\partial f/\partial \Omega_i$ through finite-differences and inverting \eqref{eq:parameter_derivatives_shape_grad}, see \citet{landremanComputingLocalSensitivity2018}. On the scale of the figure, the two shape gradients seem identical. The relative error is shown in Fig.~\ref{fig:shape_grad_iota_fom_relative_error} to be small, limited to $\sim 2\%$ at the ellipse tips, and exhibits oscillations typical of a truncated Fourier resolution. The relative error is here defined as the absolute error normalised by the $L^\infty$-norm of $\mathcal{G}_\mathrm{adjoint}$, i.e. its maximum absolute value. This choice is preferable to e.g. the $L^2$-norm, as the shape gradient and the error thereof have small average values on the boundary compared to their large values at the ellipse tips, such that unreasonably high relative errors would result at these locations if using the $L^2$-norm as normalisation. 

Furthermore, we test convergence of the shape gradient by evaluating a parameter derivative \eqref{eq:parameter_derivatives_shape_grad} for a random direction in $\Omega$. The parameter derivative is evaluated both through the adjoint shape gradient and by a forward finite-difference scheme. The relative error is shown in Fig.~\ref{fig:convergence_iota_fom} as a function of the finite-difference step size $\epsilon_\mathrm{FD}$ and the Fourier resolution, which is used both in SPEC and the Fourier-Galerkin spectral solver. As $\epsilon_\mathrm{FD}$ is reduced, the error initially decreases linearly with $\epsilon_\mathrm{FD}$, as expected from the employed forward finite-difference scheme, until it plateaus at a value governed by the finite Fourier or radial resolution. As mentioned in \S\ref{sec:basics_adjoint_methods}, errors in the adjoint shape gradient are introduced by the assumption that the constraint and adjoint PDEs are exactly satisfied. In practice, these PDEs are solved only approximately, limited by the finite Fourier and radial resolution, such that a reduction of the error with increasing resolution is to be expected.

\subsection{Quasisymmetry objective function}
\label{sec:QS_fom_shape_derivative}

For a general (non-vacuum) magnetic field with nested flux surfaces, quasisymmetry can be expressed as
\begin{equation}
    \frac{\mathbf{B}\cdot\nabla\psi\times\nabla B}{\mathbf{B}\cdot\nabla B} = -\frac{MG + NI}{N-\iota M}, \label{eq:quasisymmetry_magnetic_field_condition}
\end{equation}
where $I$ is the net toroidal plasma current and $N/M$ is the helicity of the field strength in Boozer coordinates, see e.g. \citet{helanderTheoryPlasmaConfinement2014}. For the vacuum field considered here, $I=0$. In the following, we will not consider quasi-poloidal symmetry, i.e. we will assume $M\ne 0$. If desired, it would be straight-forward to extend the derived results to include the case $M=0$.

For magnetic fields with globally nested flux surfaces labelled by $\psi$, \eqref{eq:quasisymmetry_magnetic_field_condition} is defined globally. However, we are considering a generally non-integrable field, assuming only that the boundary $\mathcal{S}$ is a flux surface. Using the generalised toroidal flux gradient defined in \eqref{eq:def_nabla_tilde_psi}, we are able to define quasisymmetry on the isolated flux surface $\mathcal{S}$, leading to the quasisymmetry (QS) objective function
\begin{equation}
    f_\mathrm{QS}(\mathcal{S}, \omega, \iota, \lambda) =  \frac{1}{2} \int_\mathcal{S} \diff S\; v_\mathrm{QS}^2(\omega, \iota, \lambda), \label{eq:definition_fQS}
\end{equation}
with
\begin{equation}
    v_\mathrm{QS} =  \mathbf{\breve{B}}\cdot\nabla \breve{B}  -  \mathbf{\breve{B}}\times\frac{\overline{\nabla\psi}}{G}\cdot\nabla \breve{B}  \left( \iota - N/M\right). \label{eq:definition_vQS}
\end{equation}
If $f_\mathrm{QS}=0$ and the field is integrable in the neighbourhood of $\mathcal{S}$, \eqref{eq:quasisymmetry_magnetic_field_condition} will be satisfied on $\mathcal{S}$, i.e. the field is quasisymmetric on the boundary.

The shape derivative of $f_\mathrm{QS}$ is derived in App.~\ref{app:derivation_QS_fom_shape_derivative}, with the final expression given in \eqref{eq:delta_fQS_tot}. Combined with the shape derivatives of $\mathcal{M}$ \eqref{eq:variation_M_tot} and $\mathcal{N}$ \eqref{eq:variation_N_tot}, the shape derivative of the Lagrangian \eqref{eq:lagrangian_general} with the quasisymmetric figure of merit follows \eqref{eq:variation_L_QS_tot}.

Requiring the Lagrangian to be stationary with respect to variations in $\iota$ and $\lambda$, the first two lines of \eqref{eq:variation_L_QS_tot} yield the adjoint equations for $q_\alpha$,
\begin{subequations}
\begin{align}
   \nabla_\Upgamma \cdot \Big( q_\alpha \mathbf{\breve{B}} \Big) = -\nabla_\Upgamma\cdot\left[ \nabla_\Upgamma\alpha \left( v_\mathrm{QS}\; \mathbf{\breve{B}}\times\frac{\overline{\nabla\psi}}{G}\cdot\nabla \breve{B}\; \frac{\iota - N/M}{\abs{\nabla_\Upgamma \alpha}^2}\right) \right] \label{eq:QS_fom_adjoint_diff_eq_qsfl}, \\
   0=\int_{\mathcal{S}} \diff S \left\{ q_\alpha \mathbf{\breve{B}} \cdot \nabla\phi + v_\mathrm{QS}\; \mathbf{\breve{B}}\times\frac{\overline{\nabla\psi}}{G}\cdot\nabla \breve{B} \left[ \frac{\nabla_\Upgamma\alpha\cdot\nabla_\Upgamma \phi}{\abs{\nabla_\Upgamma \alpha}^2} (\iota-N/M) + 1 \right] \right\} \label{eq:QS_fom_adjoint_integral_eq_qsfl}.
\end{align}
\end{subequations}
Similarly to the rotational transform objective function case, $q_\alpha$ satisfies a magnetic differential equation \eqref{eq:QS_fom_adjoint_diff_eq_qsfl} on $\mathcal{S}$, with integral condition \eqref{eq:QS_fom_adjoint_integral_eq_qsfl}. By \eqref{eq:surface_divergence_theorem}, the surface integral of \eqref{eq:QS_fom_adjoint_diff_eq_qsfl} is consistent with the magnetic field's normal component vanishing on the boundary \eqref{eq:vacuum_field_normal_BC}.

Furthermore, requiring the Lagrangian to be stationary with respect to variations in $\omega$, we obtain the adjoint equations for $q_\omega$ from the third and fourth lines of \eqref{eq:variation_L_QS_tot}
\begin{subequations}
\begin{align}
    \Updelta q_\omega & = 0   \qquad\qquad\qquad\mathrm{in}\;\mathcal{V}, \label{eq:QS_fom_adjoint_eq_qomega}\\
    \nabla q_\omega \cdot \normvec &= -\nabla_\Upgamma \cdot \Bigg\{ q_\alpha \nabla_\Upgamma \alpha + v_\mathrm{QS} \;\nabla_\Upgamma \breve{B} -  \frac{\mathbf{\breve{B}}}{\breve{B}} \nabla_\Upgamma \cdot (v_\mathrm{QS} \; \mathbf{\breve{B}})\label{eq:QS_fom_adjoint_eq_qomega_normal_BC} \\
    \nonumber & + \left(\iota-N/M\right) \left[ v_\mathrm{QS} \; \frac{\overline{\nabla\psi}}{G}\times\nabla_\Upgamma\breve{B} - \mathbf{\breve{B}}\; \nabla_\Upgamma\cdot\left(\frac{1}{\breve{B}}v_\mathrm{QS} \mathbf{\breve{B}}\times\frac{\overline{\nabla\psi}}{G}\right)  \right] \Bigg\} \quad \text{ on } \mathcal{S}.
\end{align}
\end{subequations}
Again, $q_\omega$ satisfies the Laplace equation in $\mathcal{V}$ \eqref{eq:QS_fom_adjoint_eq_qomega}, with a normal boundary condition on $\mathcal{S}$ that is the tangential divergence of a vector tangential to the surface \eqref{eq:QS_fom_adjoint_eq_qomega_normal_BC}. The boundary condition \eqref{eq:QS_fom_adjoint_eq_qomega_normal_BC} is consistent with the Laplace equation, as $\int_\mathcal{V} \diff V\; \Updelta q_\omega = \int_\mathcal{S} \diff S\; \nabla q_\omega\cdot\normvec = 0$, by \eqref{eq:surface_divergence_theorem}. 

Finally, we obtain the shape gradient from the last three lines of \eqref{eq:variation_L_QS_tot},
\begin{align}
    \mathcal{G}&_\mathrm{QS}  =  - (\normvec\cdot\nabla\breve{B}) \nabla_\Upgamma\cdot\left( v_\mathrm{QS} \mathbf{\breve{B}} \right) - v_\mathrm{QS}\;\left(\mathbf{\breve{B}}\cdot\nabla\normvec - \normvec\cdot\nabla\mathbf{\breve{B}} \right)\cdot\nabla_\Upgamma\breve{B} \label{eq:QS_fom_shape_gradient} \\
    \nonumber & + \left(\iota-N/M\right)\; \frac{|\overline{\nabla\psi}|}{G} \; \mathbf{\breve{B}}\times\nabla\breve{B}\cdot  \left[ \abs{\nabla_\Upgamma\alpha} \nabla_\Upgamma\left( \frac{v_\mathrm{QS}}{\abs{\nabla_\Upgamma\alpha}} \right) + \normvec\; v_\mathrm{QS}\; \left( \frac{\nabla_\Upgamma\alpha\cdot\nabla\normvec\cdot\nabla_\Upgamma\alpha}{\abs{\nabla_\Upgamma \alpha}^2} -h \right) \right]  \\
    \nonumber & + \mathbf{\breve{B}}\cdot\nabla q_\omega + q_\alpha \left( \normvec\cdot\nabla\mathbf{\breve{B}} - \mathbf{\breve{B}}\cdot\nabla\normvec \right)\cdot\nabla_\Upgamma\alpha + \frac{h}{2} v_\mathrm{QS}^2,
\end{align}
with $\delta\mathcal{L}_\mathrm{QS}[\delta\mathbf{x}]  = \int_{\mathcal{S}} \diff S \;(\normpert)\;\mathcal{G_\mathrm{QS}}$, and $h$ the summed curvature.

\begin{figure}
\begin{subfigure}{.5\textwidth}
  \centering
  \includegraphics[width=1\linewidth]{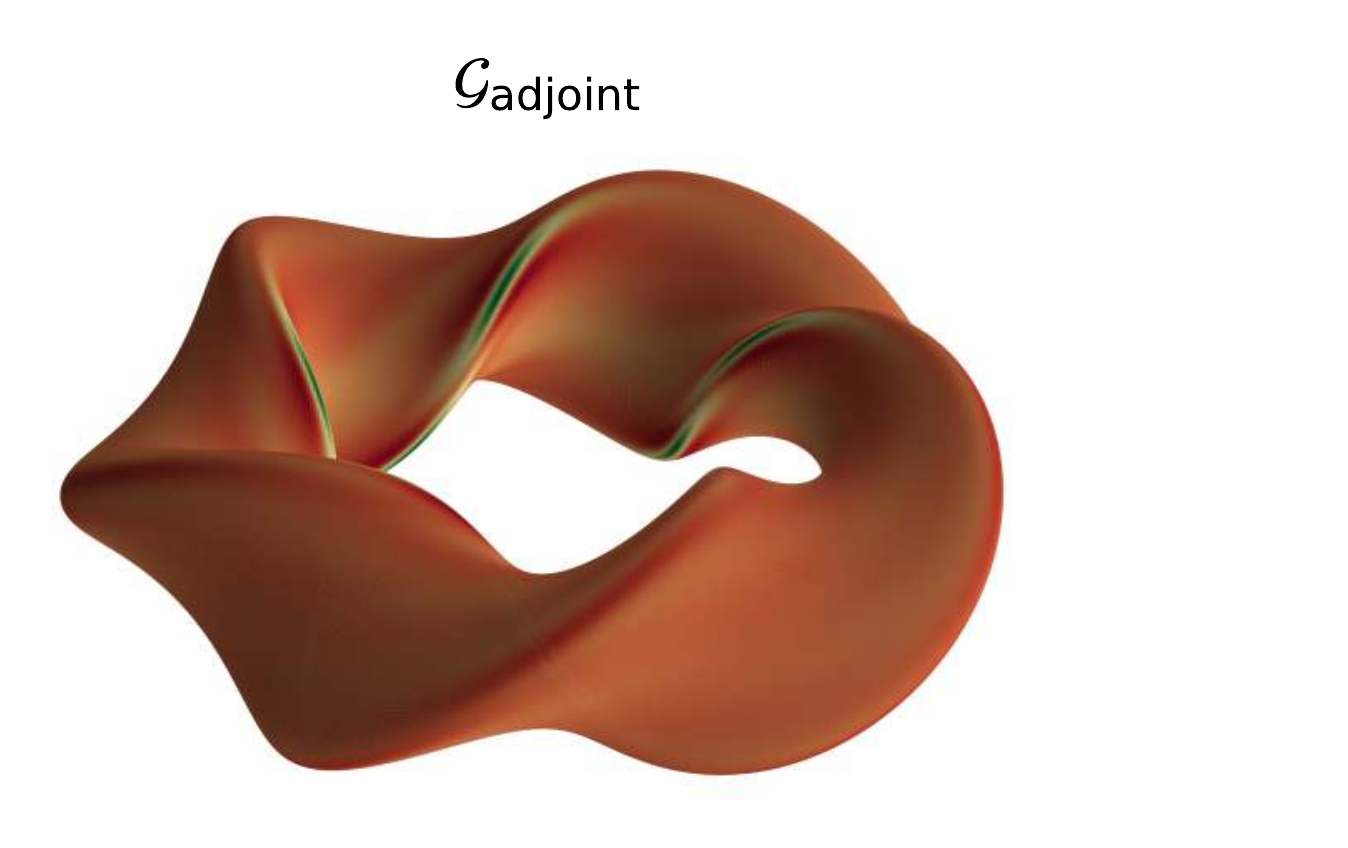}
  \caption{}
  \label{fig:shape_grad_QS_fom_adjoint}
\end{subfigure}%
\begin{subfigure}{.5\textwidth}
  \centering
  \includegraphics[width=1\linewidth]{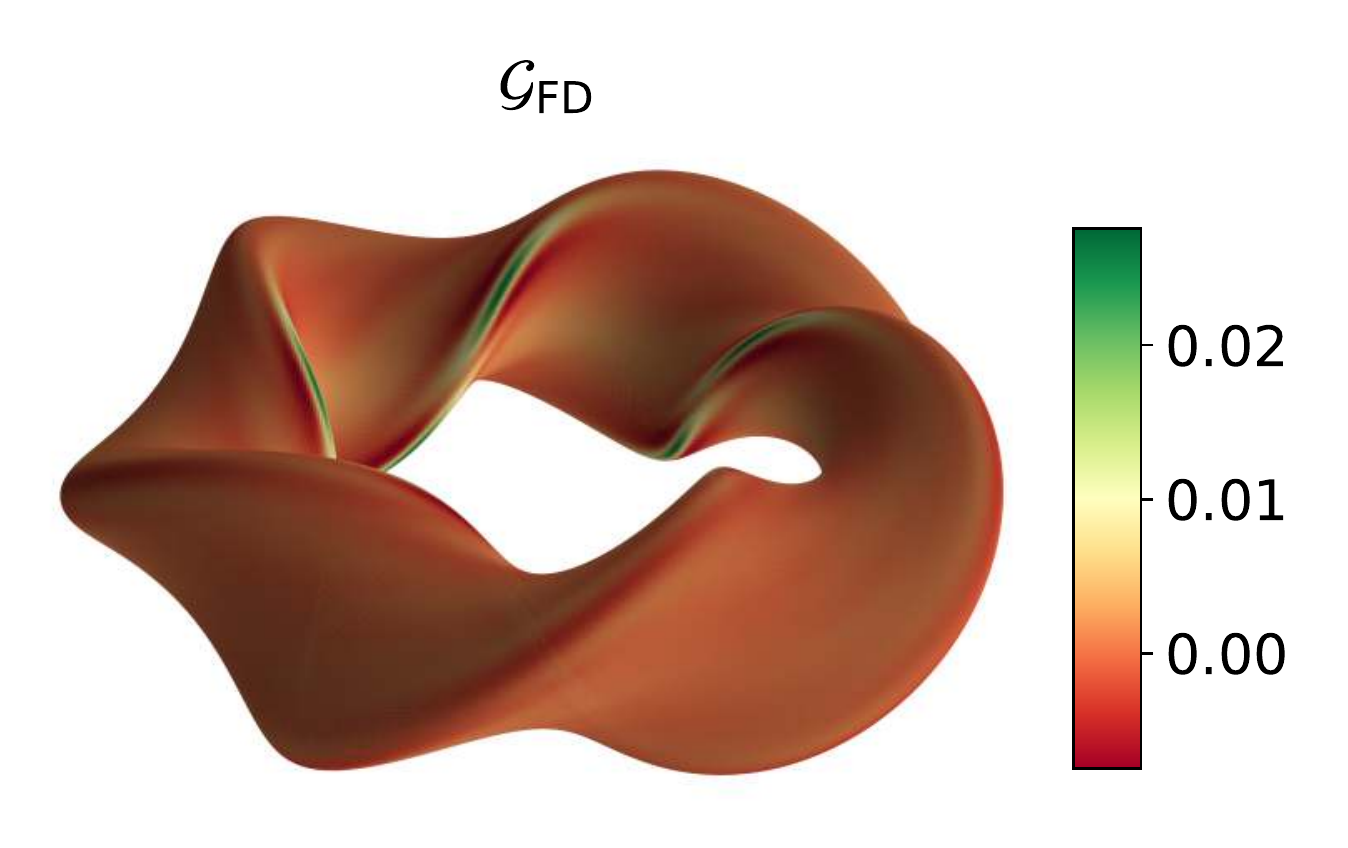}
  \caption{}
  \label{fig:shape_grad_QS_fom_findiff}
\end{subfigure}
\\
\begin{subfigure}{.5\textwidth}
  \centering
  \includegraphics[width=1\linewidth]{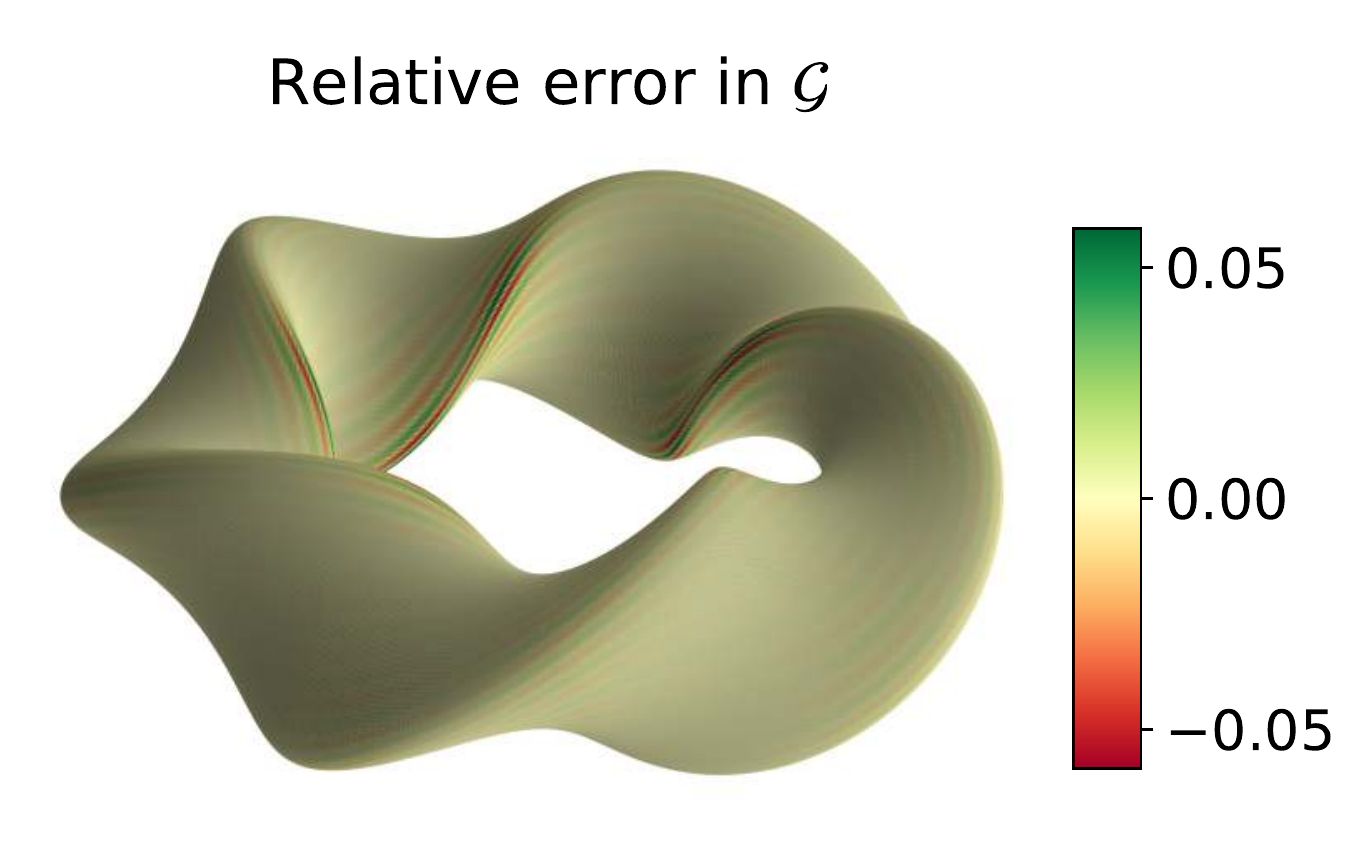}
  \caption{}
  \label{fig:shape_grad_QS_fom_relative error}
\end{subfigure}%
\begin{subfigure}{.5\textwidth}
  \centering
  \includegraphics[width=1\linewidth]{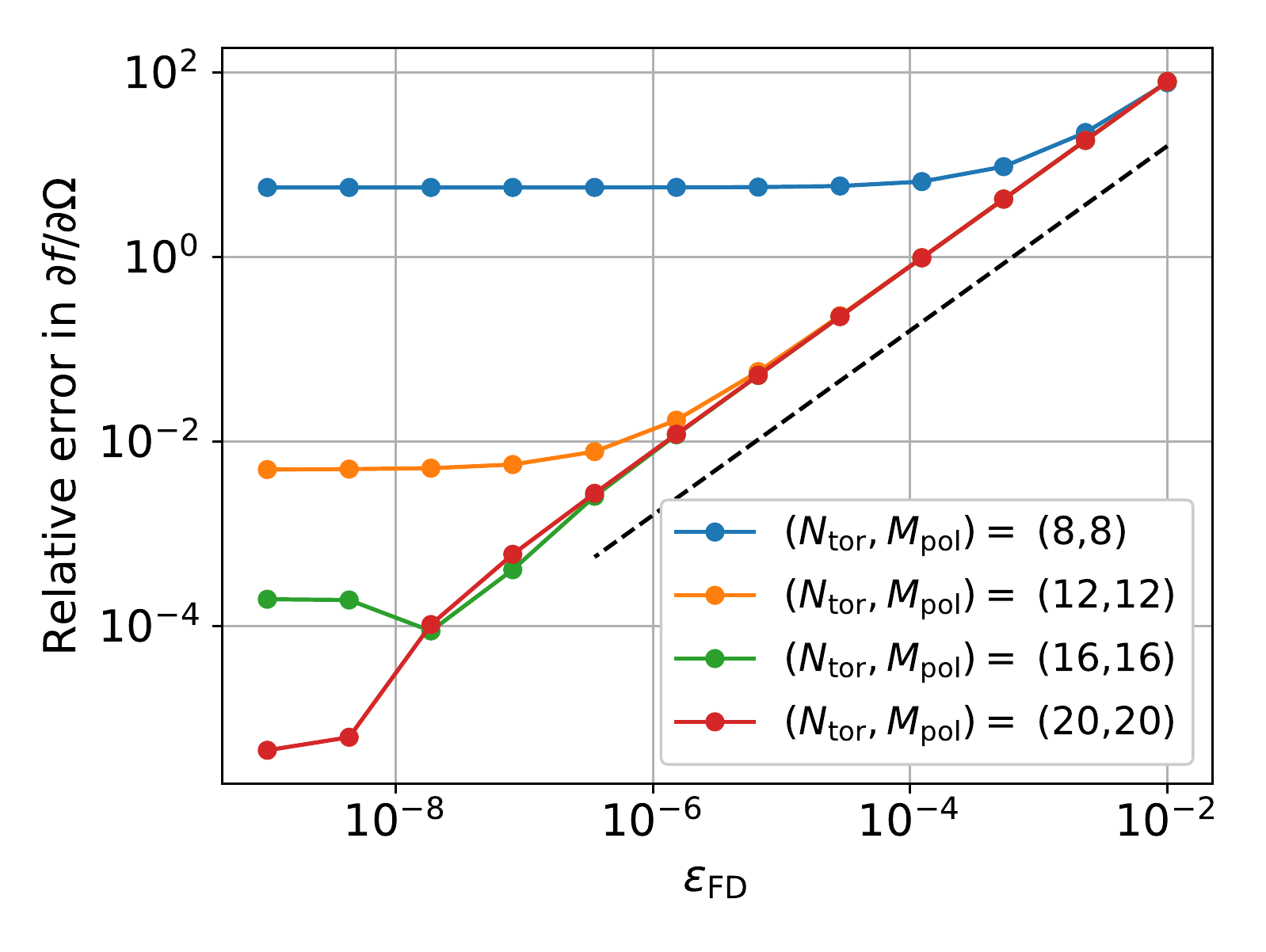}
  \caption{}
  \label{fig:convergence_QS_fom}
\end{subfigure}
\caption{Shape gradient for the quasisymmetry objective function with helicity $N/M=5$, evaluated through (a) adjoint methods and (b) a forward finite-difference scheme with step size $\epsilon_\mathrm{FD}=10^{-9}$, for the example rotating ellipse case introduced in Fig.~\ref{fig:abs_nabla_psi}, with Fourier resolution $(N_\mathrm{tor}, M_\mathrm{pol}) = (16,16)$. The relative error, defined as the absolute error normalised by the maximal absolute value of the adjoint shape gradient, is shown in (c). The convergence of the relative error in the parameter derivative \eqref{eq:parameter_derivatives_shape_grad} for a random direction in $\Omega$ is shown in (d) as a function of the step-size $\epsilon_\mathrm{FD}$ and Fourier resolution $(N_\mathrm{tor}, M_\mathrm{pol})$. The black dashed line indicates the linear scaling in $\epsilon_\mathrm{FD}$ expected from the employed forward finite-difference scheme.}
\label{fig:QS_fom_numerical_eval_and_convergence}
\end{figure}

The shape gradient \eqref{eq:QS_fom_shape_gradient} for targeted quasi-helical symmetry with helicity $N/M = 5$ is shown in Fig.~\ref{fig:shape_grad_QS_fom_adjoint} for the example rotating ellipse case introduced in Fig.~\ref{fig:abs_nabla_psi}. The shape gradient obtained through adjoint methods is verified against a finite-difference evaluation in Fig.~\ref{fig:shape_grad_QS_fom_findiff}. The error is visibly small, as is attested by the small relative error of the shape gradient shown in Fig.~\ref{fig:shape_grad_QS_fom_relative error}. Convergence of the relative error for a parameter derivative in a random direction in $\Omega$, evaluated with the adjoint method and with a centered finite-difference scheme, is shown in Fig.~\ref{fig:convergence_QS_fom}. Akin to the rotational transform figure of merit convergence study in Fig.~\ref{fig:convergence_iota_fom}, the error decreases linearly with $\epsilon_\mathrm{FD}$ until it plateaus due to finite Fourier or radial resolution. While the lowest resolution of $(N_\mathrm{tor}, M_\mathrm{pol}) = (8,8)$ seemed reasonable for the rotational transform figure of merit in Fig.~\ref{fig:convergence_iota_fom}, a higher resolution is clearly required for the quasisymmetry figure of merit. This could be due to the fact that higher derivatives of the magnetic field are involved in the shape gradient for quasisymmetry \eqref{eq:QS_fom_shape_gradient} than in the one for rotational transform \eqref{eq:shape_gradient_iota_fom}, through derivatives of $v_\mathrm{QS}$. The resulting fine-scale structure of $\mathcal{G}$ is harder to resolve with a truncated Fourier series. However, the relative errors in Figs.~\ref{fig:convergence_iota_fom}~and~\ref{fig:convergence_QS_fom} are similarly small at the highest Fourier resolutions employed.

\section{Conclusions}

In this work, we derived the adjoint equations and shape gradient for the rotational transform and quasisymmetry of a vacuum field on a surface. The shape gradients allow fast computation of derivatives with respect to the parameters that describe the geometry of the surface, which are used in optimisation and sensitivity analyses. For a boundary represented by $N$ parameters, the speed-up from the adjoint method is $O(N)$ compared to a finite-difference evaluation. 

This should enable future use of codes such as SPEC \citep{hudsonComputationMultiregionRelaxed2012} in optimisation calculations, which was hitherto neglected in favour of the more widely-used VMEC code \citep{hirshmanThreedimensionalFreeBoundary1986}. Contrary to VMEC, SPEC does not rely on the assumption of nested flux surfaces and can therefore model stochastic and island regions. In practice, employing adjoint methods and computing derivatives of quantities arising from ideal MHS equilibria is challenging, as the linearised MHS equilibrium equations possess regular singular points at every rational surface that resonates with the perturbation. These challenges can be avoided by the use of alternative equilibrium models, such as force-free magnetic fields, or the vacuum fields considered in this work. The generality of the results presented herein would also allow for their implementation in other solvers such as BIEST \citep{malhotraTaylorStatesStellarators2019}. It is left for future work to extend the vacuum field results presented herein to the more general force-free fields modeled by SPEC.  Furthermore, the adjoint methods for vacuum fields introduced in this work could be fruitfully applied to other optimisation problems, e.g. in neoclassical transport calculations.

It is generally believed that exact quasisymmetry cannot be obtained exactly in a finite volume as near-axis expansions lead to an an overdetermined system of equations \citep{garrenExistenceQuasihelicallySymmetric1991}, although that can be resolved by allowing for an anisotropic plasma pressure \citep{rodriguezSolvingProblemOverdetermination2021a, rodriguezSolvingProblemOverdetermination2021}. Exact quasisymmetry on a surface is thought generally possible \citep{garrenExistenceQuasihelicallySymmetric1991, plunkQuasiaxisymmetricMagneticFields2018}; and indeed, a vacuum solution near axisymmetry was recently found \citep{senguptaVacuumMagneticFields2021}. The shape gradient for quasisymmetry derived in this work could be used to numerically probe the existence of quasisymmetric solutions on a surface that are not close to axisymmetry. For this purpose, the shape gradient for the rotational transform objective function \eqref{eq:shape_gradient_iota_fom} could be used to avoid the axisymmetric solution at $\iota = 0$, or also to avoid low order rationals. Furthermore, the shape gradients derived herein could be used to investigate if and how optimisation for quasisymmetry and for the rotational transform compete with each other. Finally, combining the derivatives of quasisymmetry and rotational transform with previously obtained derivatives of coil shapes \citep{hudsonDifferentiatingShapeStellarator2018} and island size \citep{geraldiniAdjointMethodDetermining2021} should, in principle, allow for the efficient search of a stellarator configuration with significant rotational transform, good integrability and neoclassical confinement at the boundary, realised by simple coils.

\begin{acknowledgments}
This work was supported by U.S. DOE DE-AC02-09CH11466, DE-SC0016072 and DE-AC02–76CH03073. A.B. acknowledges the generous support of the Simons Foundation.
\end{acknowledgments}

\appendix
\section{Basics of shape differential calculus}
\label{app:basics_shape_diff_calculus}

This appendix aims to provide a brief introduction to calculus on surfaces, mainly providing useful identities required in the derivation of the adjoint equations, without strict mathematical rigour. For more details on the subject, we refer the interested reader to \citet{walkerShapesThingsPractical2015}.

In this section, we take $\mathcal{S} = \partial \mathcal{V}$ to be a closed two-dimensional surface bounding the volume $\mathcal{V}$. Let $f$ and $\mathbf{v}$ be respectively scalar and vector functions defined on $\mathcal{S}$. The extensions of these functions to a neighbourhood of $\mathcal{S}$ are denoted by $\tilde{f}$ and $\tilde{\mathbf{v}}$. Note that $f$ and $\mathbf{v}$ can also be functions defined in $\mathcal{V}$, in which case $\tilde{f}$ and $\mathbf{\tilde{v}}$ are chosen to be equal to $f$ and $\mathbf{v}$, respectively; the tangential gradient $\nabla_\Upgamma f$ and tangential divergence $\nabla_\Upgamma\cdot \mathbf{v}$ remain defined on $\mathcal{S}$.

\subsection{Differential operators on surfaces}
\label{app:diff_operators_surfaces}

The tangential gradient $\nabla_\Upgamma$ can be defined in terms of an extension as
\begin{equation}
    \nabla_\Upgamma f \equiv \nabla \tilde{f} - \normvec \left(\normvec\cdot\nabla \tilde{f} \right), \label{eq:def_tangential_gradient}
\end{equation}
where $\normvec$ is the unit normal vector on $\mathcal{S}$. The tangential gradient can thus simply be viewed as the component of the three-dimensional gradient tangential to the surface, satisfying $\normvec\cdot\nabla_\Upgamma f = 0$.

Similarly to the tangential gradient, the tangential divergence $\nabla_\Upgamma \cdot$ can be defined in terms of an extension as
\begin{equation}
    \nabla_\Upgamma\cdot \mathbf{v} = \nabla\cdot\mathbf{\tilde v} - \normvec\cdot\nabla \mathbf{\tilde v} \cdot \normvec \label{eq:def_tangential_div}.
\end{equation}
The related divergence theorem is particularly useful in our derivation of the adjoint equations,
\begin{equation}
    \int_\mathcal{S} \diff S \; \nabla_\Upgamma\cdot \mathbf{v} = \int_\mathcal{S} \diff S \; h \; \normvec\cdot\mathbf{v} \;  \label{eq:surface_divergence_theorem},
\end{equation}
where $h$ is the summed curvature, and $\mathbf{v}$ is assumed to be single-valued. In particular, it follows from \eqref{eq:surface_divergence_theorem} that $\int_\mathcal{S} \diff S \; \nabla_\Upgamma\cdot \mathbf{v}=0$ for a single-valued $\mathbf{v}$ with $\normvec\cdot\mathbf{v} = 0$.

\subsection{Transport theorems}

To evaluate the shape derivative of the Lagrangian (see \S~\ref{sec:basics_adjoint_methods}) and obtain the adjoint equations, we need expressions for the shape derivatives of volume and surface integrals. These are called transport theorems. First, for a volume functional
\begin{equation}
    J_V = \int_\mathcal{V} \diff V\; f,
\end{equation}
the shape derivative of $J_V$ is given by
\begin{equation}
    \delta J_V[\delta\mathbf{x}] = \int_\mathcal{V} \diff V\; \delta f [\delta\mathbf{x}] + \int_{\mathcal{S}} \diff S \; (\normpert) \; f. \label{eq:transport_volume}
\end{equation}
Second, for a surface functional
\begin{equation}
    J_S = \int_\mathcal{S} \diff S\; f,
\end{equation}
the shape derivative of $J_S$ is
\begin{equation}
     \delta J_S[\delta\mathbf{x}] = \int_{\mathcal{S}} \diff S \; \Big[ \delta f [\delta\mathbf{x}] + (\normpert) \left( \normvec \cdot \nabla \tilde{f} + h f \right) \Big]. \label{eq:transport_surface}
\end{equation}
In both \eqref{eq:transport_volume} and \eqref{eq:transport_surface}, the first term originates from the direct perturbation of the integrand while the second term accounts for the change of the boundary.

\subsection{Normal extension and the normal vector's shape derivative}
\label{app:normal_extension_and_normal_vector_shape_derivative}

The signed distance function $b$ is defined in a sufficiently small neighbourhood of $\mathcal{S}$ as
\begin{equation}
    b(\mathbf{r}) = \left \{ \begin{array}{l} \text{dist}(\mathbf{r},\mathcal{S}), \hspace{0.45cm} \mathbf{r} \in \mathbb{R}^3 \setminus \mathcal{V} \\
     0, \hspace{1.7cm} \mathbf{r} \in \mathcal{S} \\
     - \text{dist}(\mathbf{r},\mathcal{S}), \hspace{0.2cm} \mathbf{r} \in \mathcal{V}
     \end{array} \right . .
     \label{eq:signed_distance_function}
\end{equation}
Here, $\text{dist}(\mathbf{r},\mathcal{S})$ is the closest distance from a point $\mathbf{r}$ to the surface $\mathcal{S}$.

For quantities defined only on the surface $\mathcal{S}$, like the normal vector $\normvec$ or the field line label $\alpha$, one is free to choose an arbitrary extension to the neighbourhood of $\mathcal{S}$. Any final result (e.g. the shape gradient or adjoint equations) should be independent of this choice. A particularly convenient choice is the normal extension $\tilde{f}(\mathbf{x}) = f(\mathbf{x}-b(\mathbf{x})\nabla b(\mathbf{x}))$, as it implies $\normvec\cdot\nabla \tilde{f} = 0$ on $\mathcal{S}$. Vector functions $\mathbf{v}$ can be similarly extended.

The signed distance function can also be used to express the unit normal vector on $\mathcal{S}$ as $\normvec = \nabla b$. Let us define
\begin{equation}
    J_n = \int_\mathcal{S} \diff S\; \chi\;b = 0,
\end{equation}
with an arbitrary function $\chi$. The transport theorem \eqref{eq:transport_surface} then yields
\begin{equation}
    0 = \delta J_n[\delta\mathbf{x}] = \int_{\mathcal{S}} \diff S \; \chi\Big[ \delta b [\delta\mathbf{x}] + (\normpert) \Big],
\end{equation}
which must hold for any $\chi$, such that $\delta b[\delta\mathbf{x}] = -\normpert$. Then, the shape derivative of the normal vector follows as $\delta\normvec[\delta\mathbf{x}] = \nabla(\delta b[\delta\mathbf{x}]) = -\nabla(\normpert)$. If the unit normal vector is normally extended off $\mathcal{S}$, as is assumed in the remainder of this paper, it follows from \eqref{eq:def_tangential_gradient} that
\begin{equation}
    \delta\normvec = -\nabla_\Upgamma(\normpert). \label{eq:normal_vector_shape_derivative}
\end{equation}
Note also that the summed curvature $h$ and normal vector are related through $\nabla_\Upgamma \cdot \normvec = h$.

\section{Derivations of shape derivatives}

For ease of notation, we will in the following derivations write shape derivatives without the $[\delta\mathbf{x}]$ bracket, e.g. $\delta\omega[\delta\mathbf{x}]\rightarrow\delta\omega$, and drop tildes for extensions off the surface $\mathcal{S}$. Furthermore, we will take the field line label $\alpha$ \eqref{eq:def_field_line_label_alpha} and the normal vector $\normvec$ to be normally extended off $\mathcal{S}$, such that $\normvec\cdot\nabla\tilde{\alpha}=0$ and $\normvec\cdot\nabla\mathbf{\tilde{\hat{n}}}=0$, see App.~\ref{app:normal_extension_and_normal_vector_shape_derivative} for more details on normal extensions.

\subsection{Weak form of Laplace equation}
\label{app:derivation_laplace_eq_shape_derivative}

The weak form of the Laplace equation, previously given in \eqref{eq:weak_form_Laplace_vac}, can be partially integrated to facilitate the calculation of the shape derivative,
\begin{align}
    \mathcal{M} &= \int_{\mathcal{V}} \diff V \; q_\omega \Updelta \omega \label{eq:M_easier_delta_omega_var} = \int_{\mathcal{V}} \diff V \; \omega \Updelta q_\omega + \int_{\mathcal{S}} \diff S \; \left( q_\omega \nabla \omega - \omega \nabla q_\omega \right) \cdot \normvec \\ 
    \nonumber   &= \int_{\mathcal{V}} \diff V \; \omega \Updelta q_\omega - \int_{\mathcal{S}} \diff S \; \left( q_\omega \nabla\phi + \omega \nabla q_\omega \right) \cdot \normvec,
\end{align}
where the boundary condition on the magnetic field \eqref{eq:vacuum_field_normal_BC} was used in the last equality. Using the transport theorems \eqref{eq:transport_volume} and \eqref{eq:transport_surface}, the shape derivative of \eqref{eq:M_easier_delta_omega_var} is computed to be
\begin{align}
    \delta\mathcal{M} & = \int_{\mathcal{V}} \diff V \; \delta\omega \Updelta q_\omega + \int_{\mathcal{S}} \diff S \; \bigg\{ -  \delta \omega  \nabla q_\omega \cdot \normvec - (q_\omega \nabla\phi + \omega \nabla q_\omega) \cdot \delta\normvec \label{eq:variation_M_original} \\
    \nonumber & + (\normpert) \Big[ \omega \Updelta q_\omega - (\normvec\cdot\nabla + h) \left( q_\omega \nabla\phi\cdot\normvec + \omega \nabla q_\omega\cdot\normvec \right) \Big]\bigg\}.
\end{align}
The term involving the normal vector's shape derivative $\delta\normvec$ can be further simplified using \eqref{eq:normal_vector_shape_derivative},
\begin{align}
    & - \int_{\mathcal{S}} \diff S \; (q_\omega \nabla\phi + \omega \nabla q_\omega) \cdot \delta\normvec =  \int_{\mathcal{S}} \diff S \; \left[ - (\normpert)\; \nabla_\Upgamma\cdot  (q_\omega \nabla_\Upgamma\phi + \omega \nabla_\Upgamma q_\omega)  \right] \\
    \nonumber & = \int_{\mathcal{S}} \diff S \; (\normpert) \bigg[ h \normvec\cdot( q_\omega \nabla\phi + \omega\nabla q_\omega) + \normvec\cdot(q_\omega\nabla\nabla\phi + \omega\nabla\nabla q_\omega)\cdot\normvec -\omega \Updelta q_\omega - \nabla q_\omega \cdot \mathbf{\breve{B}}  \bigg],
\end{align}
where the surface divergence theorem \eqref{eq:surface_divergence_theorem} was used in the second equality. Inserting this expression back into \eqref{eq:variation_M_original}, the shape derivative of $\mathcal{M}$ simplifies to
\begin{equation}
    \delta\mathcal{M} = \int_{\mathcal{V}} \diff V \; \delta\omega \Updelta q_\omega - \int_{\mathcal{S}} \diff S \; \bigg[ \delta \omega  \nabla q_\omega \cdot \normvec - (\normpert)\;  \mathbf{\breve{B}}\cdot\nabla q_\omega  \bigg], \label{eq:variation_M_tot_app}
\end{equation}
where we used \eqref{eq:magnetic_field_scalar_pot}.

\subsection{Weak form of straight field line equation}
\label{app:derivation_straight_field_line_eq_shape_derivative}

The shape derivative of the straight field line equation's weak form \eqref{eq:weak_form_straight_field_line_eq} follows from the transport theorem \eqref{eq:transport_surface}, as well as the shape derivatives of the field line label $\delta\alpha = -\phi \; \delta\iota + \delta\lambda$ and the normalised magnetic field $\delta\mathbf{\breve{B}} = \nabla(\delta\omega)$,
\begin{align}
    \delta\mathcal{N} = \int_{\mathcal{S}} \diff S\; & \left[ q_\alpha  \nabla(\delta\omega) \cdot \nabla\alpha + q_\alpha \mathbf{\breve{B}} \cdot \left( \nabla(\delta\lambda) - \delta\iota \nabla\phi \right) + (\normpert)\normvec\cdot\nabla  \left( q_\alpha \mathbf{\breve{B}} \cdot \nabla\alpha \right) \right], \label{eq:variation_N_original}
\end{align}
where the summed curvature term in \eqref{eq:transport_surface} vanishes here due to the straight field line equation \eqref{eq:straight_field_line_eq}. The first term is partially integrated using \eqref{eq:surface_divergence_theorem},
\begin{equation}
    \int_{\mathcal{S}} \diff S\; q_\alpha  \nabla(\delta\omega) \cdot \nabla\alpha = \int_{\mathcal{S}} \diff S\; q_\alpha  \nabla_\Upgamma(\delta\omega) \cdot \nabla_\Upgamma \alpha  = \int_{\mathcal{S}} \diff S\; \left[ - \delta\omega \nabla_\Upgamma \cdot \left(q_\alpha \nabla_\Upgamma \alpha\right) \right],
\end{equation}
as well as the $\delta\lambda$ term,
\begin{equation}
    \int_{\mathcal{S}} \diff S\; q_\alpha \mathbf{\breve{B}} \cdot \nabla(\delta\lambda) = \int_{\mathcal{S}} \diff S\; q_\alpha \mathbf{\breve{B}} \cdot \nabla_\Upgamma(\delta\lambda)  = - \int_{\mathcal{S}} \diff S\;  \delta\lambda\; \nabla_\Upgamma \cdot \left( q_\alpha \mathbf{\breve{B}} \right).
\end{equation}
The last term in \eqref{eq:variation_N_original} can also be simplified using \eqref{eq:straight_field_line_eq},
\begin{align}
    \normvec\cdot\nabla\left( q_\alpha \mathbf{\breve{B}} \cdot \nabla\alpha \right) & = q_\alpha \left( \normvec\cdot\nabla\mathbf{\breve{B}}\cdot\nabla\alpha + \mathbf{\breve{B}}\cdot\nabla\nabla\alpha\cdot\normvec \right) = q_\alpha \left( \normvec\cdot\nabla\mathbf{\breve{B}} - \mathbf{\breve{B}}\cdot\nabla\normvec \right)\cdot\nabla_\Upgamma\alpha.
\end{align}
The shape derivative of $\mathcal{N}$ \eqref{eq:variation_N_original} finally reduces to
\begin{align}
    \delta\mathcal{N}  = \int_{\mathcal{S}}  \diff S\;  \bigg[ & - \delta\omega \;\nabla_\Upgamma \cdot \left(q_\alpha \nabla_\Upgamma \alpha\right) - \delta\iota\; q_\alpha \mathbf{\breve{B}} \cdot \nabla\phi  - \delta\lambda\; \nabla_\Upgamma \cdot \left( q_\alpha \mathbf{\breve{B}} \right) \label{eq:variation_N_tot_app}\\
    \nonumber &  + (\normpert) \; q_\alpha \left( \normvec\cdot\nabla\mathbf{\breve{B}} - \mathbf{\breve{B}}\cdot\nabla\normvec \right)\cdot\nabla_\Upgamma\alpha \bigg].
\end{align}

\subsection{Quasisymmetry figure of merit}
\label{app:derivation_QS_fom_shape_derivative}

Using the transport theorem \eqref{eq:transport_surface}, we can express the shape derivative of the quasisymmetry figure of merit \eqref{eq:definition_fQS} as
\begin{equation}
    \delta f_\mathrm{QS} = \int_{\mathcal{S}} \diff S\; \Big\{ v_\mathrm{QS}\; \delta v_\mathrm{QS} + \frac{1}{2} (\normpert)(\normvec\cdot\nabla+h)v_\mathrm{QS}^2   \Big\}. \label{eq:delta_f_QS}
\end{equation}
First, note $\delta\mathbf{\breve{B}} = \nabla(\delta\omega)$, which also gives the shape derivative of the normalised magnetic field strength as
\begin{equation}
    \delta \breve{B} = \delta \left( \sqrt{\mathbf{\breve{B}}\cdot\mathbf{\breve{B}}} \right) = \frac{\mathbf{\breve{B}}\cdot\nabla(\delta\omega)}{\breve{B}} \label{eq:delta_magnitude_B}.
\end{equation}
Furthermore, recalling $|\overline{\nabla\psi}|/G = \breve{B}/\abs{\nabla_\Upgamma\alpha}$ from \eqref{eq:def_nabla_tilde_psi}, we obtain
\begin{equation}
    \delta\left( \frac{|\overline{\nabla\psi}|}{G} \right) = \frac{|\overline{\nabla\psi}|}{G}\; \left[ \frac{\mathbf{\breve{B}}\cdot\nabla(\delta\omega)}{\breve{B}^2} - \frac{\nabla_\Upgamma\alpha\cdot \left( -\delta\iota \nabla_\Upgamma\phi + \nabla_\Upgamma (\delta\lambda) \right) }{\abs{\nabla_\Upgamma \alpha}^2} \right]. \label{eq:delta_nabla_tilde_psi}
\end{equation}
Using \eqref{eq:delta_magnitude_B}, \eqref{eq:delta_nabla_tilde_psi}, \eqref{eq:vacuum_field_pert_normal_BC} and \eqref{eq:normal_vector_shape_derivative}, the shape derivative of $v_\mathrm{QS}$ \eqref{eq:definition_vQS} can be written as
\begin{align}
    \delta v_\mathrm{QS} & = \nabla_\Upgamma(\delta\omega)\cdot\nabla_\Upgamma\breve{B} + (\normvec\cdot\nabla\breve{B})\left[ -(\normpert) \normvec\cdot\nabla\mathbf{\breve{B}}\cdot\normvec + \mathbf{\breve{B}}\cdot\nabla_\Upgamma(\normpert) \right] \\
    \nonumber & + \mathbf{\breve{B}}\cdot\nabla_\Upgamma \left( \frac{\mathbf{\breve{B}}\cdot\nabla_\Upgamma(\delta\omega)}{\breve{B}}  \right) - \delta\iota \; \mathbf{\breve{B}}\times\normvec\cdot\nabla_\Upgamma \breve{B}  \; \frac{|\overline{\nabla\psi}|}{G} -\left(\iota-N/M\right)\;\frac{|\overline{\nabla\psi}|}{G} \\
    \nonumber & \times\Bigg[ \nabla_\Upgamma(\delta\omega)\times\normvec\cdot\nabla_\Upgamma \breve{B} - \mathbf{\breve{B}}\times\nabla_\Upgamma(\normpert)\cdot\normvec (\normvec\cdot\nabla \breve{B}) + \mathbf{\breve{B}}\times\normvec\cdot\nabla_\Upgamma \left( \frac{\mathbf{\breve{B}}\cdot\nabla_\Upgamma(\delta\omega)}{\breve{B}} \right) \\
    \nonumber &  + \left( \frac{\mathbf{\breve{B}}\cdot\nabla_\Upgamma(\delta\omega)}{\breve{B}^2} - \frac{\nabla_\Upgamma\alpha\cdot \left( -\delta\iota \nabla_\Upgamma\phi + \nabla_\Upgamma (\delta\lambda) \right) }{\abs{\nabla_\Upgamma \alpha}^2} \right) \mathbf{\breve{B}}\times\normvec\cdot\nabla_\Upgamma\breve{B} \Bigg].
\end{align}
The first term in \eqref{eq:delta_f_QS} can then be partially integrated to
\begin{align}
     & \int_{\mathcal{S}} \diff S\;  v_\mathrm{QS}\; \delta v_\mathrm{QS} = \int_{\mathcal{S}} \diff S\; \Bigg\{   \delta\omega \;\nabla_\Upgamma\cdot \Bigg[ -v_\mathrm{QS} \nabla_\Upgamma \breve{B} + \frac{\mathbf{\breve{B}}}{\breve{B}} \nabla_\Upgamma\cdot(v_\mathrm{QS} \mathbf{\breve{B}})  \label{eq:vQS_delta_vQS} \\
     \nonumber & + \left(\iota-N/M\right) \Bigg( \frac{|\overline{\nabla\psi}|}{G}\; v_\mathrm{QS} \; \normvec\times\nabla_\Upgamma\breve{B} - \frac{\mathbf{\breve{B}}}{\breve{B}} \nabla_\Upgamma\cdot\bigg(\frac{|\overline{\nabla\psi}|}{G}\;v_\mathrm{QS} \mathbf{\breve{B}}\times\normvec\bigg) \\
     \nonumber &  + v_\mathrm{QS}\frac{|\overline{\nabla\psi}|}{G} \frac{\mathbf{\breve{B}}}{\tilde{B^2}}  \mathbf{\breve{B}}\times\normvec\cdot\nabla_\Upgamma\breve{B} \Bigg) \Bigg] - \delta\iota \; v_\mathrm{QS}\frac{|\overline{\nabla\psi}|}{G} \mathbf{\breve{B}}\times\normvec\cdot\nabla_\Upgamma\breve{B}  \left[ \left(\iota-N/M\right) \frac{\nabla_\Upgamma\alpha\cdot\nabla_\Upgamma\phi}{\abs{\nabla_\Upgamma\alpha}^2} + 1\right]  \\
     \nonumber & - \delta\lambda \;\left(\iota-N/M\right)\;\nabla_\Upgamma\cdot\Bigg[ v_\mathrm{QS} \frac{|\overline{\nabla\psi}|}{G} \frac{\nabla_\Upgamma\alpha}{\abs{\nabla_\Upgamma\alpha}^2} \; \mathbf{\breve{B}}\times\normvec\cdot\nabla_\Upgamma\breve{B} \Bigg] \\
     \nonumber & - (\normpert)\Bigg[  \mathbf{\breve{B}}\cdot \nabla_\Upgamma\left( v_\mathrm{QS}\;\normvec\cdot\nabla\breve{B} \right) - \left(\iota-N/M\right)\; \nabla_\Upgamma\cdot\Bigg( v_\mathrm{QS} \frac{|\overline{\nabla\psi}|}{G} \mathbf{\breve{B}}\times(\nabla\breve{B} -\nabla_\Upgamma\breve{B} ) \Bigg) \Bigg] \Bigg\},
\end{align}
with repeated use of the surface divergence theorem \eqref{eq:surface_divergence_theorem}, and using $\nabla_\Upgamma\cdot\mathbf{\breve{B}} = -\normvec\cdot\nabla\mathbf{\breve{B}}\cdot\normvec$ from the magnetic field being divergence-less and \eqref{eq:def_tangential_div}. The terms in the last line can be simplified, first using
\begin{align}
     \mathbf{\breve{B}}\cdot \nabla_\Upgamma\left( v_\mathrm{QS}\;\normvec\cdot\nabla\breve{B} \right) = (\normvec\cdot\nabla\breve{B})  \mathbf{\breve{B}}\cdot\nabla v_\mathrm{QS} + v_\mathrm{QS}\left( \mathbf{\breve{B}}\cdot\nabla\normvec\cdot\nabla\breve{B} +  \mathbf{\breve{B}}\cdot\nabla\nabla\breve{B}\cdot\normvec \right) \label{eq:normpert_term_vQS_delta_vQS_1}.
\end{align}
Furthermore, note that
\begin{align}
    \nabla_\Upgamma\cdot(\mathbf{\breve{B}}\times\nabla\breve{B}) = - \normvec\cdot\nabla(\mathbf{\breve{B}}\times\nabla\breve{B})\cdot\normvec,
\end{align}
where we used \eqref{eq:def_tangential_div}, $\nabla\times\mathbf{\breve{B}} = 0$ and $\nabla\times\nabla\breve{B}=0$; and
\begin{align}
    \mathbf{\breve{B}}\times\nabla\breve{B}\cdot\nabla_\Upgamma \Bigg(\frac{|\overline{\nabla\psi}|}{G}\Bigg) &  =  -\frac{|\overline{\nabla\psi}|}{G}\mathbf{\breve{B}}\times\nabla\breve{B}\cdot\left( \frac{1}{\breve{B}} \normvec(\normvec\cdot\nabla\breve{B}) + \frac{1}{\abs{\nabla_\Upgamma\alpha}}   \nabla_\Upgamma\abs{\nabla_\Upgamma\alpha} \right),
\end{align}
using \eqref{eq:def_tangential_gradient}. The terms of the last parenthesis in \eqref{eq:vQS_delta_vQS} now simplify to
\begin{align}
     & \nabla_\Upgamma\cdot\Bigg( v_\mathrm{QS}\; \frac{|\overline{\nabla\psi}|}{G}\; \mathbf{\breve{B}}\times\nabla\breve{B} \Bigg) \\ 
     \nonumber & = \frac{|\overline{\nabla\psi}|}{G} \bigg[ \mathbf{\breve{B}}\times\nabla\breve{B}\cdot\left( \abs{\nabla_\Upgamma\alpha} \nabla_\Upgamma \left( \frac{v_\mathrm{QS}}{\abs{\nabla_\Upgamma\alpha}}\right) - v_\mathrm{QS}\;\normvec \frac{\normvec\cdot\nabla\breve{B}}{\breve{B}}  \right) - v_\mathrm{QS}\; \normvec\cdot\nabla(\mathbf{\breve{B}}\times\nabla\breve{B})\cdot\normvec  \bigg],
\end{align}
and
\begin{equation}
    \nabla_\Upgamma \cdot \Bigg( v_\mathrm{QS}\; \frac{|\overline{\nabla\psi}|}{G}\; \mathbf{\breve{B}}\times \nabla_\Upgamma\breve{B}   \Bigg) =  h\; v_\mathrm{QS}\; \frac{|\overline{\nabla\psi}|}{G}\; \normvec\cdot \mathbf{\breve{B}}\times \nabla_\Upgamma\breve{B}  ,
\end{equation}
where we used the fact that $\mathbf{\breve{B}}\times \nabla_\Upgamma\breve{B}$ is normal to the surface, and $\nabla_\Upgamma\cdot\normvec = h$. 

We now turn to the normal derivative of $v_\mathrm{QS}$, the second term in \eqref{eq:delta_f_QS}. Using the fact that $\normvec$ and $\alpha$ are normally extended, we can write
\begin{equation}
    \normvec\cdot\nabla(\mathbf{\breve{B}}\times\normvec\cdot\nabla\breve{B})  = -\normvec\cdot\nabla(\mathbf{\breve{B}}\times\nabla\breve{B})\cdot\normvec
\end{equation}
and
\begin{equation}
    \normvec\cdot\nabla\abs{\nabla_\Upgamma\alpha} = \frac{\normvec\cdot\nabla\nabla_\Upgamma\alpha\cdot\nabla_\Upgamma\alpha}{\abs{\nabla_\Upgamma\alpha}} =\frac{\normvec\cdot\nabla\nabla\alpha\cdot\nabla_\Upgamma\alpha}{\abs{\nabla_\Upgamma\alpha}} = -\frac{\nabla_\Upgamma\alpha\cdot\nabla\normvec\cdot\nabla_\Upgamma\normvec}{\abs{\nabla_\Upgamma\alpha}},
\end{equation}
such that
\begin{equation}
    \normvec\cdot\nabla\left( \frac{|\overline{\nabla\psi}|}{G} \right) = \frac{|\overline{\nabla\psi}|}{G} \left[ \frac{\normvec\cdot\nabla\breve{B}}{\breve{B}} + \frac{ \nabla_\Upgamma\alpha\cdot\nabla\normvec\cdot\nabla_\Upgamma\alpha}{\abs{\nabla_\Upgamma\alpha}^2} \right].
\end{equation}
This allows us to express the normal derivative of $v_\mathrm{QS}$ as
\begin{align}
    &\normvec \cdot \nabla v_\mathrm{QS} = \; \normvec\cdot\nabla\mathbf{\breve{B}}\cdot\nabla\breve{B} + \normvec\cdot\nabla\nabla\breve{B}\cdot\mathbf{\breve{B}} \label{eq:normal_derivative_vQS} \\
    \nonumber & - (\iota-N/M) \frac{|\overline{\nabla\psi}|}{G} \left[  -\normvec\cdot\nabla(\mathbf{\breve{B}}\times\nabla\breve{B})\cdot\normvec + \left( \frac{\normvec\cdot\nabla\breve{B}}{\breve{B}} + \frac{ \nabla_\Upgamma\alpha\cdot\nabla\normvec\cdot\nabla_\Upgamma\alpha}{\abs{\nabla_\Upgamma\alpha}^2} \right) \mathbf{\breve{B}}\times\normvec\cdot\nabla\breve{B} \right].
\end{align}
Combining \eqref{eq:delta_f_QS}, \eqref{eq:vQS_delta_vQS} (and simplifications thereafter) and \eqref{eq:normal_derivative_vQS}, the shape derivative of $f_\mathrm{QS}$ can finally be expressed as
\begin{align}
    \delta & f_\mathrm{QS} = \int_{\mathcal{S}} \diff S\; \Bigg\{   \delta\omega \;\nabla_\Upgamma\cdot \Bigg[ -v_\mathrm{QS} \nabla_\Upgamma \breve{B} + \frac{\mathbf{\breve{B}}}{\breve{B}} \nabla_\Upgamma\cdot(v_\mathrm{QS} \mathbf{\breve{B}})  \label{eq:delta_fQS_tot} \\
     \nonumber & + \left(\iota-N/M\right) \Bigg(  v_\mathrm{QS} \; \frac{\overline{\nabla\psi}}{G}\times\nabla_\Upgamma\breve{B} - \mathbf{\breve{B}}\; \nabla_\Upgamma\cdot\left(\frac{1}{\breve{B}} v_\mathrm{QS} \mathbf{\breve{B}}\times\frac{\overline{\nabla\psi}}{G}\right) \Bigg) \Bigg] \\
     \nonumber & - \delta\iota \; v_\mathrm{QS} \mathbf{\breve{B}}\times\frac{\overline{\nabla\psi}}{G}\cdot\nabla_\Upgamma\breve{B}  \left[ \left(\iota-N/M\right) \frac{\nabla_\Upgamma\alpha\cdot\nabla_\Upgamma\phi}{\abs{\nabla_\Upgamma\alpha}^2} + 1\right]  \\
     \nonumber & - \delta\lambda \;\left(\iota-N/M\right)\;\nabla_\Upgamma\cdot\left[ v_\mathrm{QS}  \frac{\nabla_\Upgamma\alpha}{\abs{\nabla_\Upgamma\alpha}^2} \; \mathbf{\breve{B}}\times\frac{\overline{\nabla\psi}}{G}\cdot\nabla_\Upgamma\breve{B} \right] \\
     \nonumber & + (\normpert)\; \Bigg[ \left(\iota-N/M\right)\; \frac{|\overline{\nabla\psi}|}{G} \; \mathbf{\breve{B}}\times\nabla\breve{B}\cdot \left( \nabla_\Upgamma v_\mathrm{QS} - v_\mathrm{QS} \frac{\nabla_\Upgamma \abs{\nabla_\Upgamma\alpha}}{\abs{\nabla_\Upgamma\alpha}} \right)\\ 
    \nonumber & + \frac{h}{2} v_\mathrm{QS}^2  - (\normvec\cdot\nabla\breve{B}) \nabla_\Upgamma\cdot\left( v_\mathrm{QS} \mathbf{\breve{B}} \right) - v_\mathrm{QS}\;\left(\mathbf{\breve{B}}\cdot\nabla\normvec - \normvec\cdot\nabla\mathbf{\breve{B}} \right)\cdot\nabla_\Upgamma\breve{B}   \\
    \nonumber & - v_\mathrm{QS} \; \mathbf{\breve{B}}\times\frac{\overline{\nabla\psi}}{G}\cdot\nabla \breve{B}  \;  \left( \iota - N/M\right) \left( \frac{\nabla_\Upgamma\alpha\cdot\nabla\normvec\cdot\nabla_\Upgamma\alpha}{\abs{\nabla_\Upgamma \alpha}^2} -h \right) \Bigg] \Bigg\},
\end{align}
where we used
\begin{align}
    \mathbf{\breve{B}}\cdot\nabla\normvec\cdot\nabla\breve{B}  -  \normvec\cdot\nabla\mathbf{\breve{B}}\cdot\nabla\breve{B} = \left(\mathbf{\breve{B}}\cdot\nabla\normvec - \normvec\cdot\nabla\mathbf{\breve{B}} \right)\cdot\nabla_\Upgamma\breve{B} +\nabla_\Upgamma \cdot \mathbf{\breve{B}}\;(\normvec\cdot\nabla\breve{B}).
\end{align}

Combining \eqref{eq:variation_M_tot_app}, \eqref{eq:variation_N_tot_app} and \eqref{eq:delta_fQS_tot}, and rearranging, the shape derivative of the Lagrangian for the quasisymmetric figure of merit can be written as
\begin{align}
    \delta & \mathcal{L}_\mathrm{QS} =  -\int_{\mathcal{S}} \diff S\; \delta\lambda \; \nabla_\Upgamma\cdot\left[ q_\alpha\mathbf{\breve{B}} + \left(\iota-N/M\right) v_\mathrm{QS} \frac{\nabla_\Upgamma\alpha}{\abs{\nabla_\Upgamma\alpha}^2} \; \mathbf{\breve{B}}\times\frac{\overline{\nabla\psi}}{G}\cdot\nabla_\Upgamma\breve{B} \right] \label{eq:variation_L_QS_tot}\\
    \nonumber & -\delta\iota\;  \int_{\mathcal{S}} \diff S \Bigg\{ q_\alpha \mathbf{\breve{B}} \cdot \nabla\phi + v_\mathrm{QS} \mathbf{\breve{B}}\times\frac{\overline{\nabla\psi}}{G}\cdot\nabla_\Upgamma\breve{B}  \left[ \left(\iota-N/M\right) \frac{\nabla_\Upgamma\alpha\cdot\nabla_\Upgamma\phi}{\abs{\nabla_\Upgamma\alpha}^2} + 1\right] \Bigg\} \\
    \nonumber & + \int_{\mathcal{V}} \diff V \; \delta\omega \; \Updelta q_\omega - \int_{\mathcal{S}} \diff S\; \delta\omega \Bigg\{ \normvec\cdot\nabla q_\omega + \nabla_\Upgamma\cdot \Bigg[ -v_\mathrm{QS} \nabla_\Upgamma \breve{B} + \frac{\mathbf{\breve{B}}}{\breve{B}} \nabla_\Upgamma\cdot(v_\mathrm{QS} \mathbf{\breve{B}})  \\
     \nonumber & + \left(\iota-N/M\right) \Bigg(  v_\mathrm{QS} \; \frac{\overline{\nabla\psi}}{G}\times\nabla_\Upgamma\breve{B} - \mathbf{\breve{B}}\; \nabla_\Upgamma\cdot\left(\frac{1}{\breve{B}} v_\mathrm{QS} \mathbf{\breve{B}}\times\frac{\overline{\nabla\psi}}{G}\right) \Bigg) \Bigg] \Bigg\} \\
     \nonumber & + \int_{\mathcal{S}} \diff S \; (\normpert)\; \Bigg\{  - (\normvec\cdot\nabla\breve{B}) \nabla_\Upgamma\cdot\left( v_\mathrm{QS} \mathbf{\breve{B}} \right) - v_\mathrm{QS}\;\left(\mathbf{\breve{B}}\cdot\nabla\normvec - \normvec\cdot\nabla\mathbf{\breve{B}} \right)\cdot\nabla_\Upgamma\breve{B} \\
    \nonumber & + \left(\iota-N/M\right) \frac{|\overline{\nabla\psi}|}{G}  \mathbf{\breve{B}}\times\nabla\breve{B}\cdot  \left[ \abs{\nabla_\Upgamma\alpha} \nabla_\Upgamma\left( \frac{v_\mathrm{QS}}{\abs{\nabla_\Upgamma\alpha}} \right) + \normvec\; v_\mathrm{QS} \left( \frac{\nabla_\Upgamma\alpha\cdot\nabla\normvec\cdot\nabla_\Upgamma\alpha}{\abs{\nabla_\Upgamma \alpha}^2} -h \right) \right]  \\
    \nonumber & + \mathbf{\breve{B}}\cdot\nabla q_\omega + q_\alpha \left( \normvec\cdot\nabla\mathbf{\breve{B}} - \mathbf{\breve{B}}\cdot\nabla\normvec \right)\cdot\nabla_\Upgamma\alpha + \frac{h}{2} v_\mathrm{QS}^2, \Bigg\}.
\end{align}
The shape derivative of the Lagrangian directly provides the adjoint equations for $q_\omega$ and $q_\alpha$, as well as the shape gradient, as shown in \S\ref{sec:QS_fom_shape_derivative}.

Competing interests: The authors declare none.

 \bibliographystyle{jpp}

 \bibliography{references}

\begin{thebibliography}{42}
\expandafter\ifx\csname natexlab\endcsname\relax\def\natexlab#1{#1}\fi
\def\au#1{#1} \def\ed#1{#1} \def\yr#1{#1}\def\at#1{#1}\def\jt#1{\textit{#1}}
  \def\bt#1{#1}\def\bvol#1{\textbf{#1}} \def\vol#1{#1} \def\pg#1{#1}
  \def\publ#1{#1}\def\arxiv#1{#1}\def\org#1{#1}\def\st#1{\textit{#1}}

\bibitem[Anderson {\em et~al.\/}(1995)Anderson, Almagri, Anderson, Matthews,
  Talmadge \& Shohet]{andersonHelicallySymmetricExperiment1995}
{\sc \au{Anderson, F. S.~B.}, \au{Almagri, A.~F.}, \au{Anderson, D.~T.},
  \au{Matthews, P.~G.}, \au{Talmadge, J.~N.} \& \au{Shohet, J.~L.}} \yr{1995}
  \at{The {Helically} {Symmetric} {Experiment}, ({HSX}) {Goals}, {Design} and
  {Status}}.  \jt{Fusion Technology}  \bvol{27}~(3T),  \pg{273--277}.

\bibitem[Antonsen {\em et~al.\/}(2019)Antonsen, Paul \&
  Landreman]{antonsenAdjointApproachCalculating2019}
{\sc \au{Antonsen, T.}, \au{Paul, E.~J.} \& \au{Landreman, M.}} \yr{2019}
  \at{Adjoint approach to calculating shape gradients for three-dimensional
  magnetic confinement equilibria}.  \jt{Journal of Plasma Physics}
  \bvol{85}~(2).

\bibitem[Bader {\em et~al.\/}(2019)Bader, Drevlak, Anderson, Faber, Hegna,
  Likin, Schmitt \& Talmadge]{baderStellaratorEquilibriaReactor2019}
{\sc \au{Bader, A.}, \au{Drevlak, M.}, \au{Anderson, D.~T.}, \au{Faber, B.~J.},
  \au{Hegna, C.~C.}, \au{Likin, K.~M.}, \au{Schmitt, J.~C.} \& \au{Talmadge,
  J.~N.}} \yr{2019}  \at{Stellarator equilibria with reactor relevant energetic
  particle losses}.  \jt{Journal of Plasma Physics}  \bvol{85}~(5),
  \pg{905850508}.

\bibitem[Beidler {\em et~al.\/}(1990)Beidler, Grieger, Herrnegger, Harmeyer,
  Kisslinger, Lotz, Maassberg, Merkel, Nührenberg, Rau, Sapper, Sardei,
  Scardovelli, Schlüter \& Wobig]{beidlerPhysicsEngineeringDesign1990}
{\sc \au{Beidler, C.}, \au{Grieger, G.}, \au{Herrnegger, F.}, \au{Harmeyer,
  E.}, \au{Kisslinger, J.}, \au{Lotz, W.}, \au{Maassberg, H.}, \au{Merkel, P.},
  \au{Nührenberg, J.}, \au{Rau, F.}, \au{Sapper, J.}, \au{Sardei, F.},
  \au{Scardovelli, R.}, \au{Schlüter, A.} \& \au{Wobig, H.}} \yr{1990}
  \at{Physics and {Engineering} {Design} for {Wendelstein} {VII}-{X}}.
  \jt{Fusion Technology}  \bvol{17}~(1),  \pg{148--168}.

\bibitem[Boozer(2019)]{boozerCurlfreeMagneticFields2019}
{\sc \au{Boozer, A.~H.}} \yr{2019}  \at{Curl-free magnetic fields for
  stellarator optimization}.  \jt{Physics of Plasmas}  \bvol{26}~(10),
  \pg{102504}.

\bibitem[Burby {\em et~al.\/}(2020)Burby, Kallinikos \&
  MacKay]{burbyMathematicsQuasisymmetry2020}
{\sc \au{Burby, J.~W.}, \au{Kallinikos, N.} \& \au{MacKay, R.~S.}} \yr{2020}
  \at{Some mathematics for quasi-symmetry}.  \jt{Journal of Mathematical
  Physics}  \bvol{61}~(9),  \pg{093503}.

\bibitem[Delfour \& Zolésio(2011)]{delfourShapesGeometries2011}
{\sc \au{Delfour, M.~C.} \& \au{Zolésio, J.~P.}} \yr{2011} {\em Shapes and
  {Geometries}\/}. {\em Advances in {Design} and {Control}\/} .  \publ{Society
  for Industrial and Applied Mathematics}.

\bibitem[Drevlak {\em et~al.\/}(2013)Drevlak, Brochard, Helander, Kisslinger,
  Mikhailov, Nührenberg, Nührenberg \&
  Turkin]{drevlakESTELLQuasiToroidallySymmetric2013}
{\sc \au{Drevlak, M.}, \au{Brochard, F.}, \au{Helander, P.}, \au{Kisslinger,
  J.}, \au{Mikhailov, M.}, \au{Nührenberg, C.}, \au{Nührenberg, J.} \&
  \au{Turkin, Y.}} \yr{2013}  \at{{ESTELL}: {A} {Quasi}-{Toroidally}
  {Symmetric} {Stellarator}}.  \jt{Contributions to Plasma Physics}
  \bvol{53}~(6),  \pg{459--468}.

\bibitem[Garren \& Boozer(1991)]{garrenExistenceQuasihelicallySymmetric1991}
{\sc \au{Garren, D.~A.} \& \au{Boozer, A.~H.}} \yr{1991}  \at{Existence of
  quasihelically symmetric stellarators}.  \jt{Physics of Fluids B: Plasma
  Physics}  \bvol{3}~(10),  \pg{2822--2834}.

\bibitem[Geraldini {\em et~al.\/}(2021)Geraldini, Landreman \&
  Paul]{geraldiniAdjointMethodDetermining2021}
{\sc \au{Geraldini, A.}, \au{Landreman, M.} \& \au{Paul, E.}} \yr{2021}  \at{An
  adjoint method for determining the sensitivity of island size to magnetic
  field variations}.  \jt{Journal of Plasma Physics}  \bvol{87}~(3),
  \pg{905870302}.

\bibitem[Giuliani {\em et~al.\/}(2020)Giuliani, Wechsung, Cerfon, Stadler \&
  Landreman]{giulianiSinglestageGradientbasedStellarator2020}
{\sc \au{Giuliani, A.}, \au{Wechsung, F.}, \au{Cerfon, A.}, \au{Stadler, G.} \&
  \au{Landreman, M.}} \yr{2020}  \at{Single-stage gradient-based stellarator
  coil design: {Optimization} for near-axis quasi-symmetry}.
  \jt{arXiv:2010.02033 [physics]} ArXiv: 2010.02033.

\bibitem[Hall \&
  McNamara(1975)]{hallThreedimensionalEquilibriumAnisotropic1975}
{\sc \au{Hall, L.~S.} \& \au{McNamara, B.}} \yr{1975}  \at{Three-dimensional
  equilibrium of the anisotropic, finite-pressure guiding-center plasma:
  {Theory} of the magnetic plasma}.  \jt{Physics of Fluids}  \bvol{18}~(5),
  \pg{552}.

\bibitem[Helander(2014)]{helanderTheoryPlasmaConfinement2014}
{\sc \au{Helander, P.}} \yr{2014}  \at{Theory of plasma confinement in
  non-axisymmetric magnetic fields}.  \jt{Reports on Progress in Physics}
  \bvol{77}~(8),  \pg{087001}, publisher: IOP Publishing.

\bibitem[Henneberg {\em et~al.\/}(2019)Henneberg, Drevlak, Nührenberg,
  Beidler, Turkin, Loizu \&
  Helander]{hennebergPropertiesNewQuasiaxisymmetric2019}
{\sc \au{Henneberg, S.}, \au{Drevlak, M.}, \au{Nührenberg, C.}, \au{Beidler,
  C.}, \au{Turkin, Y.}, \au{Loizu, J.} \& \au{Helander, P.}} \yr{2019}
  \at{Properties of a new quasi-axisymmetric configuration}.  \jt{Nuclear
  Fusion}  \bvol{59}~(2),  \pg{026014}.

\bibitem[Henneberg {\em et~al.\/}(2020)Henneberg, Drevlak \&
  Helander]{hennebergImprovingFastparticleConfinement2020}
{\sc \au{Henneberg, S.~A.}, \au{Drevlak, M.} \& \au{Helander, P.}} \yr{2020}
  \at{Improving fast-particle confinement in quasi-axisymmetric stellarator
  optimization}.  \jt{Plasma Physics and Controlled Fusion}  \bvol{62}~(1),
  \pg{014023}.

\bibitem[Hirshman {\em et~al.\/}(1986)Hirshman, van RIJ \&
  Merkel]{hirshmanThreedimensionalFreeBoundary1986}
{\sc \au{Hirshman, S.}, \au{van RIJ, W.} \& \au{Merkel, P.}} \yr{1986}
  \at{Three-dimensional free boundary calculations using a spectral {Green}'s
  function method}.  \jt{Computer Physics Communications}  \bvol{43}~(1),
  \pg{143--155}.

\bibitem[Hudson {\em et~al.\/}(2018)Hudson, Zhu, Pfefferlé \&
  Gunderson]{hudsonDifferentiatingShapeStellarator2018}
{\sc \au{Hudson, S.}, \au{Zhu, C.}, \au{Pfefferlé, D.} \& \au{Gunderson, L.}}
  \yr{2018}  \at{Differentiating the shape of stellarator coils with respect to
  the plasma boundary}.  \jt{Physics Letters A}  \bvol{382}~(38),
  \pg{2732--2737}.

\bibitem[Hudson {\em et~al.\/}(2012)Hudson, Dewar, Dennis, Hole, McGann, von
  Nessi \& Lazerson]{hudsonComputationMultiregionRelaxed2012}
{\sc \au{Hudson, S.~R.}, \au{Dewar, R.~L.}, \au{Dennis, G.}, \au{Hole, M.~J.},
  \au{McGann, M.}, \au{von Nessi, G.} \& \au{Lazerson, S.}} \yr{2012}
  \at{Computation of multi-region relaxed magnetohydrodynamic equilibria}.
  \jt{Physics of Plasmas}  \bvol{19}~(11),  \pg{112502}.

\bibitem[Ku \& Boozer(2011)]{kuNewClassesQuasihelically2011}
{\sc \au{Ku, L.} \& \au{Boozer, A.}} \yr{2011}  \at{New classes of
  quasi-helically symmetric stellarators}.  \jt{Nuclear Fusion}  \bvol{51}~(1),
   \pg{013004}.

\bibitem[Landreman {\em et~al.\/}(2021)Landreman, Medasani \&
  Zhu]{landremanStellaratorOptimizationGood2021}
{\sc \au{Landreman, M.}, \au{Medasani, B.} \& \au{Zhu, C.}} \yr{2021}
  \at{Stellarator optimization for good magnetic surfaces at the same time as
  quasisymmetry}.  \jt{arXiv:2106.14930 [physics]} ArXiv: 2106.14930.

\bibitem[Landreman \& Paul(2018)]{landremanComputingLocalSensitivity2018}
{\sc \au{Landreman, M.} \& \au{Paul, E.}} \yr{2018}  \at{Computing local
  sensitivity and tolerances for stellarator physics properties using shape
  gradients}.  \jt{Nuclear Fusion}  \bvol{58}~(7),  \pg{076023}.

\bibitem[Landreman \& Paul(2021)]{landremanMagneticFieldsPrecise2021}
{\sc \au{Landreman, M.} \& \au{Paul, E.}} \yr{2021}  \at{Magnetic fields with
  precise quasisymmetry}.  \jt{arXiv:2108.03711 [physics]} ArXiv: 2108.03711.

\bibitem[Malhotra {\em et~al.\/}(2019)Malhotra, Cerfon, Imbert-Gérard \&
  O'Neil]{malhotraTaylorStatesStellarators2019}
{\sc \au{Malhotra, D.}, \au{Cerfon, A.}, \au{Imbert-Gérard, L.-M.} \&
  \au{O'Neil, M.}} \yr{2019}  \at{Taylor states in stellarators: {A} fast
  high-order boundary integral solver}.  \jt{Journal of Computational Physics}
  \bvol{397},  \pg{108791}.

\bibitem[Mercier(1964)]{mercierEquilibriumStabilityToroidal1964}
{\sc \au{Mercier, C.}} \yr{1964}  \at{Equilibrium and stability of a toroidal
  magnetohydrodynamic system in the neighbourhood of a magnetic axis}.
  \jt{Nuclear Fusion}  \bvol{4}~(3),  \pg{213--226}.

\bibitem[Nies(2021)]{niesDataCodesPaper2021}
{\sc \au{Nies, R.}} \yr{2021} Dataset from: "{Adjoint} methods for
  quasisymmetry of vacuum fields on a surface". DOI: 10.5281/zenodo.5248498.

\bibitem[Nührenberg \&
  Zille(1988)]{nuhrenbergQuasihelicallySymmetricToroidal1988}
{\sc \au{Nührenberg, J.} \& \au{Zille, R.}} \yr{1988}  \at{Quasi-helically
  symmetric toroidal stellarators}.  \jt{Physics Letters A}  \bvol{129}~(2),
  \pg{113--117}.

\bibitem[Paul(2020)]{paulAdjointMethodsStellarator2020}
{\sc \au{Paul, E.}} \yr{2020}  \at{Adjoint methods for stellarator shape
  optimization and sensitivity analysis}.  \jt{arXiv:2005.07633 [physics]} .

\bibitem[Paul {\em et~al.\/}(2018)Paul, Landreman, Bader \&
  Dorland]{paulAdjointMethodGradientbased2018}
{\sc \au{Paul, E.}, \au{Landreman, M.}, \au{Bader, A.} \& \au{Dorland, W.}}
  \yr{2018}  \at{An adjoint method for gradient-based optimization of
  stellarator coil shapes}.  \jt{Nuclear Fusion}  \bvol{58}~(7),  \pg{076015}.

\bibitem[Paul {\em et~al.\/}(2019)Paul, Abel, Landreman \&
  Dorland]{paulAdjointMethodNeoclassical2019}
{\sc \au{Paul, E.~J.}, \au{Abel, I.~G.}, \au{Landreman, M.} \& \au{Dorland,
  W.}} \yr{2019}  \at{An adjoint method for neoclassical stellarator
  optimization}.  \jt{Journal of Plasma Physics}  \bvol{85}~(5).

\bibitem[Paul {\em et~al.\/}(2020)Paul, Antonsen, Landreman \&
  Cooper]{paulAdjointApproachCalculating2020}
{\sc \au{Paul, E.~J.}, \au{Antonsen, T.}, \au{Landreman, M.} \& \au{Cooper,
  W.~A.}} \yr{2020}  \at{Adjoint approach to calculating shape gradients for
  three-dimensional magnetic confinement equilibria. {Part} 2. {Applications}}.
   \jt{Journal of Plasma Physics}  \bvol{86}~(1).

\bibitem[Paul {\em et~al.\/}(2021)Paul, Landreman \&
  Antonsen]{paulGradientbasedOptimization3D2021}
{\sc \au{Paul, E.~J.}, \au{Landreman, M.} \& \au{Antonsen, T.}} \yr{2021}
  \at{Gradient-based optimization of {3D} {MHD} equilibria}.  \jt{Journal of
  Plasma Physics}  \bvol{87}~(2).

\bibitem[Plunk \& Helander(2018)]{plunkQuasiaxisymmetricMagneticFields2018}
{\sc \au{Plunk, G.~G.} \& \au{Helander, P.}} \yr{2018}  \at{Quasi-axisymmetric
  magnetic fields: weakly non-axisymmetric case in a vacuum}.  \jt{Journal of
  Plasma Physics}  \bvol{84}~(2),  \pg{905840205}.

\bibitem[Qu {\em et~al.\/}(2020)Qu, Pfefferlé, Hudson, Baillod, Kumar, Dewar
  \& Hole]{quCoordinateParameterisationSpectral2020}
{\sc \au{Qu, Z.~S.}, \au{Pfefferlé, D.}, \au{Hudson, S.~R.}, \au{Baillod, A.},
  \au{Kumar, A.}, \au{Dewar, R.~L.} \& \au{Hole, M.~J.}} \yr{2020}
  \at{Coordinate parameterisation and spectral method optimisation for
  {Beltrami} field solver in stellarator geometry}.  \jt{Plasma Physics and
  Controlled Fusion}  \bvol{62}~(12),  \pg{124004}.

\bibitem[Rodríguez \& Bhattacharjee(2021{\natexlab{{\em
  a\/}}})]{rodriguezSolvingProblemOverdetermination2021a}
{\sc \au{Rodríguez, E.} \& \au{Bhattacharjee, A.}} \yr{2021{\natexlab{{\em
  a\/}}}}  \at{Solving the problem of overdetermination of quasisymmetric
  equilibrium solutions by near-axis expansions. {I}. {Generalized} force
  balance}.  \jt{Physics of Plasmas}  \bvol{28}~(1),  \pg{012508}, publisher:
  American Institute of Physics.

\bibitem[Rodríguez \& Bhattacharjee(2021{\natexlab{{\em
  b\/}}})]{rodriguezSolvingProblemOverdetermination2021}
{\sc \au{Rodríguez, E.} \& \au{Bhattacharjee, A.}} \yr{2021{\natexlab{{\em
  b\/}}}}  \at{Solving the problem of overdetermination of quasisymmetric
  equilibrium solutions by near-axis expansions. {II}. {Circular} axis
  stellarator solutions}.  \jt{Physics of Plasmas}  \bvol{28}~(1),
  \pg{012509}, publisher: American Institute of Physics.

\bibitem[Rodríguez {\em et~al.\/}(2020)Rodríguez, Helander \&
  Bhattacharjee]{rodriguezNecessarySufficientConditions2020}
{\sc \au{Rodríguez, E.}, \au{Helander, P.} \& \au{Bhattacharjee, A.}}
  \yr{2020}  \at{Necessary and sufficient conditions for quasisymmetry}.
  \jt{Physics of Plasmas}  \bvol{27}~(6),  \pg{062501}.

\bibitem[Sengupta {\em et~al.\/}(2021)Sengupta, Paul, Weitzner \&
  Bhattacharjee]{senguptaVacuumMagneticFields2021}
{\sc \au{Sengupta, W.}, \au{Paul, E.~J.}, \au{Weitzner, H.} \&
  \au{Bhattacharjee, A.}} \yr{2021}  \at{Vacuum magnetic fields with exact
  quasisymmetry near a flux surface. {Part} 1. {Solutions} near an axisymmetric
  surface}.  \jt{Journal of Plasma Physics}  \bvol{87}~(2),  \pg{905870205}.

\bibitem[Sokołowski \&
  Zolésio(1992)]{sokolowskiIntroductionShapeOptimization1992}
{\sc \au{Sokołowski, J.} \& \au{Zolésio, J.~P.}} \yr{1992} {\em Introduction
  to shape optimization: shape sensitivity analysis\/}. {\em Springer series in
  computational mathematics\/} 16.  \publ{Berlin ; New York: Springer-Verlag}.

\bibitem[Spitzer(1958)]{spitzerStellaratorConcept1958}
{\sc \au{Spitzer, L.}} \yr{1958}  \at{The {Stellarator} {Concept}}.
  \jt{Physics of Fluids}  \bvol{1}~(4),  \pg{253}.

\bibitem[Walker(2015)]{walkerShapesThingsPractical2015}
{\sc \au{Walker, S.~W.}} \yr{2015} {\em The shapes of things: a practical guide
  to differential geometry and the shape derivative\/}. {\em Advances in design
  and control\/} .  \publ{Philadelphia: Society for Industrial and Applied
  Mathematics}.

\bibitem[Zarnstorff {\em et~al.\/}(2001)Zarnstorff, Berry, Brooks, Fredrickson,
  Fu, Hirshman, Hudson, Ku, Lazarus, Mikkelsen, Monticello, Neilson, Pomphrey,
  Reiman, Spong, Strickler, Boozer, Cooper, Goldston, Hatcher, Isaev, Kessel,
  Lewandowski, Lyon, Merkel, Mynick, Nelson, Nuehrenberg, Redi, Reiersen,
  Rutherford, Sanchez, Schmidt \& White]{zarnstorffPhysicsCompactAdvanced2001}
{\sc \au{Zarnstorff, M.~C.}, \au{Berry, L.~A.}, \au{Brooks, A.},
  \au{Fredrickson, E.}, \au{Fu, G.-Y.}, \au{Hirshman, S.}, \au{Hudson, S.},
  \au{Ku, L.-P.}, \au{Lazarus, E.}, \au{Mikkelsen, D.}, \au{Monticello, D.},
  \au{Neilson, G.~H.}, \au{Pomphrey, N.}, \au{Reiman, A.}, \au{Spong, D.},
  \au{Strickler, D.}, \au{Boozer, A.}, \au{Cooper, W.~A.}, \au{Goldston, R.},
  \au{Hatcher, R.}, \au{Isaev, M.}, \au{Kessel, C.}, \au{Lewandowski, J.},
  \au{Lyon, J.~F.}, \au{Merkel, P.}, \au{Mynick, H.}, \au{Nelson, B.~E.},
  \au{Nuehrenberg, C.}, \au{Redi, M.}, \au{Reiersen, W.}, \au{Rutherford, P.},
  \au{Sanchez, R.}, \au{Schmidt, J.} \& \au{White, R.~B.}} \yr{2001}
  \at{Physics of the compact advanced stellarator {NCSX}}.  \jt{Plasma Physics
  and Controlled Fusion}  \bvol{43}~(12A),  \pg{A237--A249}.

\bibitem[Zhu {\em et~al.\/}(2018)Zhu, Hudson, Song \&
  Wan]{zhuNewMethodDesign2018}
{\sc \au{Zhu, C.}, \au{Hudson, S.~R.}, \au{Song, Y.} \& \au{Wan, Y.}} \yr{2018}
   \at{New method to design stellarator coils without the winding surface}.
  \jt{Nuclear Fusion}  \bvol{58}~(1),  \pg{016008}.

\end{thebibliography}

\end{document}